\documentclass[preprint,12pt]{elsarticle}

\usepackage[margin=3.0cm]{geometry}
\usepackage{diagbox}
\usepackage{tabularray}
\usepackage{array}
\usepackage{caption}
\usepackage{dirtytalk}
\usepackage{amsfonts} 
\usepackage{multirow}
\usepackage{subcaption}
\usepackage{booktabs,tabularx,graphicx}
\usepackage{amsmath}
\usepackage{stackengine}
\usepackage{comment}
\usepackage{changepage}

\usepackage[font=footnotesize,labelfont=bf, textfont=sc]{caption}
\usepackage{lipsum}
\usepackage{algorithmic}

\usepackage{rotating}
\usepackage{wasysym}
 
\usepackage{colortbl}
\usepackage{hhline}

\journal{}

\begin{document}

\begin{frontmatter}

\title{A Hybrid Intelligent System  for Protection of Transmission Lines Connected to PV Farms based on Linear Trends}

\author[a]{Pallav Kumar Bera
    }
\author[b]{Samita Rani Pani}
\author[c]{Can Isik}
\author[d]{Ramesh C. Bansal}

\address[a]{Electrical Engineering, Western Kentucky University, Bowling Green, USA}
\address[b]{School of Electrical Engineering, KIIT University, Bhubaneswar, Odisha, India}
\address[c]{EECS Department, Syracuse University, Syracuse, NY, USA}
\address[d]{EE Department, University of Sharjah, Sharjah, UAE}

\begin{abstract} 
Conventional relays face challenges for transmission lines connected to inverter-based resources (IBRs). In this article, a single-ended intelligent protection of the transmission line in the zone between the grid and the PV farm is suggested. The method employs a fuzzy logic and random forest (RF)-based hybrid system to detect faults based on combined linear trend attributes of the 3-phase currents. The fault location is determined and the faulty phase is detected. RF feature selection is used to obtain the optimal linear trend feature. The performance of the methodology is examined for abnormal events such as faults, capacitor and load-switching operations simulated in PSCAD/EMTDC 
on IEEE 9-bus system obtained by varying various fault and switching parameters.
Additionally, when validating the suggested strategy, consideration is given to the effects of conditions such as the presence of double circuit lines, PV capacity, sampling rate, data window length, noise, high impedance faults, CT saturation, compensation devices, evolving and cross-country faults, and far-end and near-end faults. The findings indicate that the suggested strategy can be used to deal with a variety of system configurations and situations while still safeguarding such complex power transmission networks.

\end{abstract}


\end{frontmatter}

\section{INTRODUCTION}

The generation of electricity using renewable energy sources (RESs) has radically grown in recent years. This trend is expected to grow as governments, businesses, and individuals around the world recognize the environmental and economic advantages of transitioning away from fossil fuels to cleaner energy sources. The predominant portion of RESs — solar photovoltaic (PV), Type-III wind farm (WF), and Type-IV WF often integrate into the grid through high-voltage transmission lines (t-lines) to transmit power generated at remote sites through a power electronic converter. 
Integration of RESs has changed the topology of existing power systems having bidirectional power flow and different fault current levels \cite{bansal}.
The fault ride-through (FRT) requirements set by modern grid codes and the intermittent nature of these inverter-based resources (IBRs) control the characteristics of fault currents \cite{singh2018}.
Hence, unlike synchronous generators, the fault current depends on the control approach, inverter control parameters, and power-system fault conditions.  
This increased penetration of IBR sources like solar and wind power into the electrical grid poses challenges to the effectiveness of existing protection schemes, including distance, differential, and directional protection.

In \cite{sikander}, the study explores how the integration of large-scale PV plants affects the performance of traditional distance relaying-based t-line protection systems.
Distance relay underreaches due to less contribution of fault current from PV plants with power electronics \cite{Banaiemoqadam2020}. The PV side distance relays may mal-operate due to unique fault characteristics of PV plants \cite{liang}. 
In \cite{ritwik}, it is demonstrated how the phase distance elements in zones 1 and 2 exhibit overreach and underreach, respectively, due to oscillating apparent impedance resulting from currents injected by the IBRs.
The integration of PV systems affecting the reliability of phase and ground distance elements supervised by negative-sequence directional elements is discussed again in \cite{kou2020}.
Haddadi et al. \cite{haddadi2021} described the concerns with t-line protection and detailed the use of negative-sequence values for identifying unbalanced faults.
In \cite{pvqiang}, it is investigated how large-scale grid connection of PV lowers the dependability of differential protection due to the disparity in short circuit behavior between PV inverters and conventional synchronous machines. Yang et al. \cite{yang} describes difficulties with sensitivity in t-line differential protection.
In \cite{chen}, an actual event is examined, and issues with an overcurrent relay's directionality are noted. The proficiency of existing protection relays to determine the proper direction of the fault was impacted by the distortion of the fault signals and the change in angular disparities between voltages and currents. The altered fault current seen by overcurrent relays affects fault localization and causes false tripping and blinding of protection \cite{kumar}. 
In \cite{hosiyar2}, the impedance characteristics of IBRs with FRT under European and North American grid codes are described. It is shown that the impedances seen are different from the actual fault impedances due to the control strategies implemented. Moreover, the already challenging task of detecting high impedance faults becomes even more formidable with the integration of PV \cite{kavi}.
Given that short circuit studies for these IBRs are not standardized and are instead specific to their designs,
it highlights the need for an alternative and inclusive protection system to support conventional protective relays.

The intricate fault current properties of IBRs have been the subject of numerous research studies in the existing literature.
In \cite{fang2019}, an enhanced scheme utilizing delay and zero sequence impedance was devised
to prevent the malfunctioning of traditional distance protection for IBRs due to amplitude and phase offset in measured impedance. In \cite{pallavauto}, autoregressive coefficients of the 3-phase currents were used for fault detection and classification for a t-line connected to large scale WF.
Saber et al. \cite{saber22} proposed a differential protection technique reliant on the phase current samples' signs on both ends of the t-line connected to WF. A transient current signal-based current differential protection approach is explored in \cite{JIA2018}, which uses correlation focusing on the similarity of the waveforms and the polarity of transient signals from line ends. A distance protection system based on positive sequence networks independent of resistance and plant parameters for the t-lines connected to the PV plants is proposed in \cite{GHORBANI2023}. A directional relaying scheme relying on positive sequence components of the fault and pre-fault voltages and currents for the t-line connected with the PV plant is proposed in \cite{AKTER2022}.

The Power System Relaying and Control Committee's Oct. 2023 report emphasizes that with increased availability of massive volumes of high-fidelity sampled data in temporal and frequency domains, the development and implementation of machine learning (ML)-based solutions for challenging protection tasks could be successful [21].
Research articles have also suggested leveraging machine learning (ML) systems to enhance fault detection and classification operation.
The work presented in \cite{omar2018} introduces an intelligent protection technique which utilizes an adaptive neuro-fuzzy inference system to detect, locate, and classify fault types occurring in large-scale grid-connected wind farms (WFs).
A fault identification method based on positive-sequence currents, coupled with an empirical mode decomposition (EMD) accompanied random forest (RF) is suggested for a TCSC compensated line in \cite{sauviktcsc21}. However, the above articles use very few features and it's improbable that a single feature will be able to capture every fault characteristic concerning PVs. Therefore, before applying any learning technique, it's vital to assess different features and utilize a feature selection process. 
The effect of PV-fed t-lines in infeed conditions on distance protection is analyzed and an impedance calculation method using SVM is proposed for transmission systems with PV integration \cite{pvsvm}. An ML technique to detect and classify faults in t-lines connected to PV and WFs is proposed in \cite{mdpi_pvwf}. However, various scenarios are not taken into account in these works, including high impedance faults, double circuit line faults, evolving and cross-country faults, CT saturation, and so on. Also, protection systems that rely on communication and synchronized data to detect faults have limitations in the event of communication failures \cite{CHOWDHURY}.

In this work, a novel single-ended combined linear trend (CLT) based intelligent protection technique is developed for the protection of lines connected to PV farms taking into account multiple power system scenarios and a variety of features.
The study's originality and main contributions are summed up as follows: 
\noindent
\begin{itemize}
   \item It proposes a fuzzy logic and RF-based hybrid intelligent method to detect and locate faults. 
   The technique is tested against various conditions simulated by taking into account several parameters that may affect the fault currents.
  \item The CLT-based protection is verified for the instances of noise, near-end and far-end faults, high impedance faults, double circuit lines, CT saturation, cross-country and evolving faults, TCSC, different sampling frequencies, window sizes, and change in PV capacity and test system.
  \item  Phasor estimation and sequence component analysis are not required for the suggested approach.
  \item Being single-ended, it is fast, without any data loss or synchronization errors, and there is no need for a communication medium.
  \item The fault and transient dataset having more than 28000 cases is uploaded to IEEE Dataport\cite{datapv}.
\end{itemize}

The remaining portions of the manuscript are arranged as follows: In section II the modeling of the PV system and simulation of faults, load-switching, and capacitor-switching events in the IEEE 9-bus test system is performed. The CLT-based hybrid intelligent protection strategy, feature selection, CLT feature, and RF classifier are elaborated in section III. Section IV exhibits the findings of fault detection, localization, and phase selection for the PV in the IEEE 9-bus system. The impact of noise, sampling rate, window size, CT saturation, number of PV units, double circuit lines, TCSC, evolving and cross-country faults, high impedance faults, near-end and far-end faults, and change in test system are explored in section V. Comparative analysis with various recent articles is provided in Section VI. A summary of the observations can be found in Section VII. 

 \begin{figure}[ht]
	\centering
	\captionsetup{justification=centering,textfont=small}
	\includegraphics[width=4.5 in, height=1.95 in]{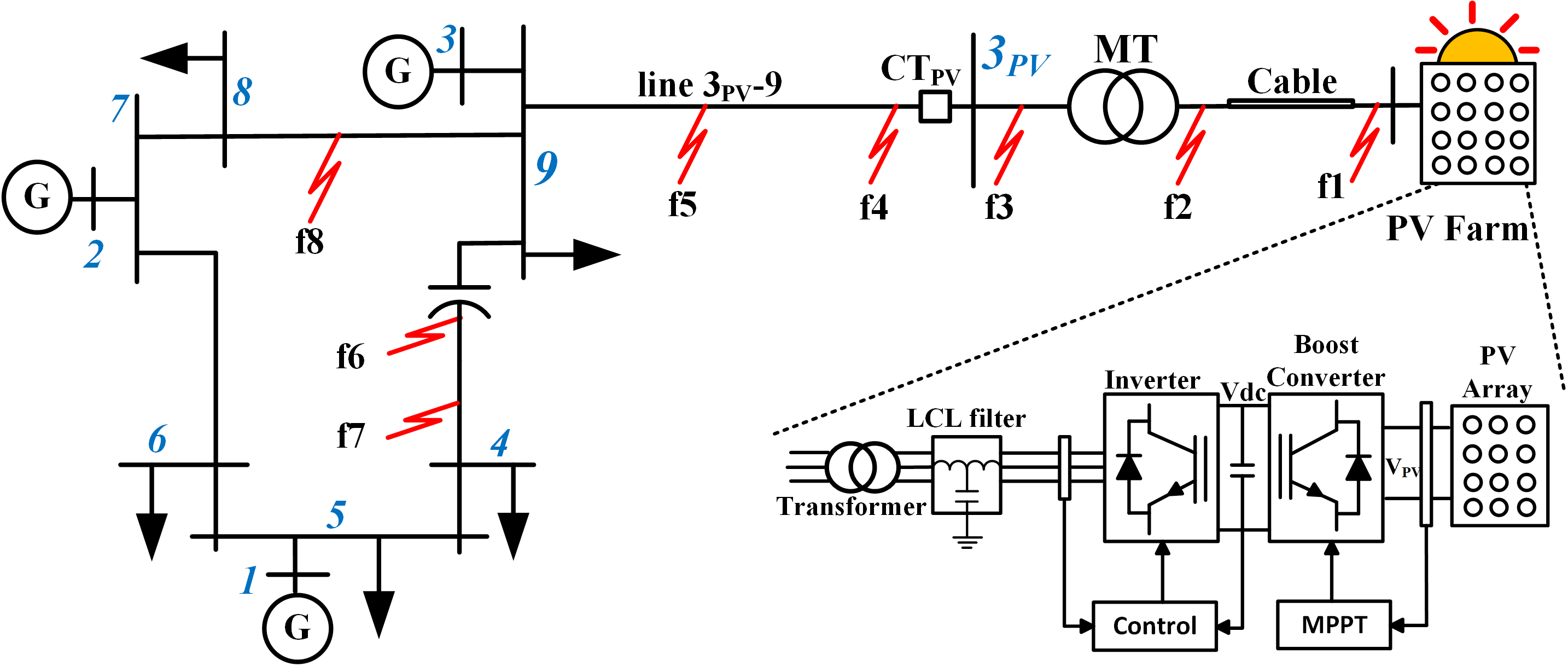}
	\caption{IEEE 9-bus with PV Plant at bus-9.}
	\label{9busPV}
\end{figure}

\begin{table}[ht]
	\centering
	\renewcommand{\arraystretch}{0.95}
	\setlength{\tabcolsep}{4 pt}
	\caption{Specifications of the various system components}
	\label{system_parameters}
	\footnotesize
	\begin{tabular}{|l|l|l|} 
		\hline
		\rowcolor[rgb]{0.91,0.91,0.91}  Component                                       & Parameters                     & Value                  \\ 
		\hline
		\multirow{5}{*}{PV}                                                            & capacity (single unit)       & 0.25MW                    \\
		& number of units                 & 400                    \\
		& dc voltage \& capacitance & 1.16kV,10000$\mu$F      \\
		&  LCL filter                    &   5mH, 39.24$\mu$F, 5mH                  \\
		&  Damper                    &    19.62$\mu$F, 10.7$\Omega$, 2.5mH                  \\
		& rated voltage  \& frequency       & 0.65kV, 60 Hz           \\ 
		\hline
		\multirow{3}{*}{t-line}                                                    & positive seq. impedance    & 0.95 + 29.90j$\Omega$    \\
		& zero seq. impedance        & 32.8 + 109.8j$\Omega$  \\
		& length  \&      voltage          & 100km, 230kV           \\ 
		\hline
		\multirow{2}{*}{collector system cable}                                                  & positive seq. impedance    & 0.18 + 0.23j$\Omega$     \\
		& zero seq. impedance        & 0.23 + 0.15j$\Omega$     \\  \hline
		\multirow{3}{*}{\begin{tabular}[c]{@{}l@{}}main\\transformer (MT)\\(after cable)\end{tabular}} & rated power                 & 300MVA                 \\
		& transformation ratio                     & 33kV/230kV             \\
		& connection                     & YNYN                    \\  \hline
		\multirow{3}{*}{\begin{tabular}[c]{@{}l@{}}transformer\\(before cable)\end{tabular}}     & rated power                 & 250kVA                 \\
		& transformation ratio                     & 33kV/0.65kV      \\
		& connection                     & YNd  \\\hline              
	\end{tabular}
\end{table}

\begin{table}
	\renewcommand{\arraystretch}{1.1}
	\setlength{\tabcolsep}{6 pt}
	\centering
	\caption{{Parameters for fault Simulation} }\label{parameters1}
	\footnotesize
	\begin{tabular}{|l|l|} 
		\hline
		\multicolumn{2}{|c|}{{\cellcolor[rgb]{0.89,0.89,0.89}}\textbf{Fault Events}}                                                            \\ 
		
		\hline
		Fault location           & f1, f2, f3, f4, f5, f6, f7, f8     (8)                                                                                \\ 
		
		\hline
		Fault resistance & 0.01, 1, 10 $\Omega$ (3)                                                                                      \\ 
		\hline
		Fault inception angle    & 0$^\circ$, 60$^\circ$, 120$^\circ$, 180$^\circ$, 240$^\circ$, 300$^\circ$ (6)                                 \\ 
		\hline
		Fault type               & $ag, ab, ac, abg, acg, abcg, bg, bcg, bc, cg$ (10)                                                            \\ 
		\hline
		Priority         & P and Q (2)                                                                                            
		\\ 
		
		\hline
		\multicolumn{2}{|l|}{{\cellcolor[rgb]{0.949,0.949,0.949}}Total fault cases =$ 8\times 3\times 6 \times 10 \times 2$ = 2880}    \\ 
		\hline
	\end{tabular}
\end{table}

\begin{table}
	\renewcommand{\arraystretch}{1.1}
	\setlength{\tabcolsep}{6 pt}
	
	\caption{{Simulation parameters \& values for other transients.} }\label{parameters2}
	\centering
	\footnotesize
	\begin{tabular}{|l|l|} 
		\hline
		\multicolumn{2}{|c|}{{\cellcolor[rgb]{0.89,0.89,0.89}}\textbf{Non-Fault Events }}                                                        \\ 
		\hline
		Switching angle          & 0$^\circ$ to 360$^\circ$ in steps of 15$^\circ$ (25)                                                          \\ 
		\hline
		Generator (bus-3)     & Disconnected/Connected (2)                                                                                   \\ 
		\hline
		Location                 & Bus-4, 8, 9    (3)                                                                                            \\ 
		\hline
		Load/Capacitor Rating    & 4                                                                                                            \\ 
		
		\hline
		Priority         & P and Q (2)                                                                                              \\

		\hline
		\multicolumn{2}{|l|}{Capacitor-switching events = $25\times 2\times 3 \times 4 \times 2$ = 1200}                                          \\ 
		\hline
		\multicolumn{2}{|l|}{Load-switching events = $25\times 2\times 3 \times 4 \times 2$ = 1200}                                               \\ 
		\hline
		\multicolumn{2}{|l|}{{\cellcolor[rgb]{0.949,0.949,0.949}}Total non-fault events = 2400}                                                   \\
		\hline
	\end{tabular}
\end{table}

\begin{figure}[ht]
	\centering
	\includegraphics [width=3.8 in, height= 1.6 in] {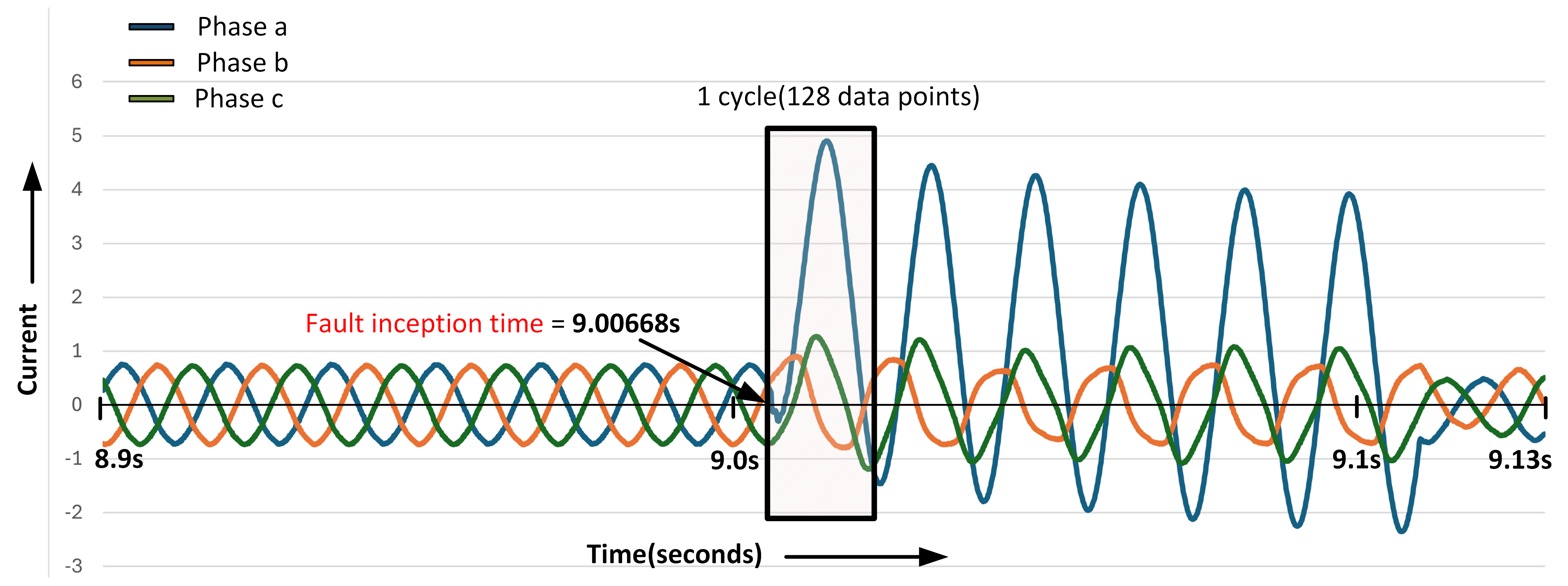} 
	\caption{3 phase fault current for lg fault at location f1}\label{lg fault}
\end{figure}

\section{System Modeling and Simulation of Transients}

The fault and other transient scenarios are simulated using the IEEE 9-bus test setup (Fig. \ref{9busPV}), which has the PV attached to bus-9. Table \ref{system_parameters} provides information about the system implemented in PSCAD/EMTDC. The PV farm consists of a power plant controller (PPC), PV array, boost converter, DC-AC converter, and scaling component. PPC determines the PV farm's reference active and reactive powers based on the measured and reference values. Temperature and irradiance both affect how much power a PV array produces.
The boost converter tracks the maximum power point or regulates the DC voltage. The DC-AC inverter is managed so that the PV farm's dynamics can be realized through a variety of control scenarios. Several inverter units in the PV farm are modeled using the scaling component. Low-frequency oscillations are dampened using a damper and a filter is used to reduce harmonic effects on the grid \cite{pscad}. The t-line $3_{PV}$-9 under consideration is 100km in length with positive sequence impedance of (0.898 + 28.3j)$\times$$10^{-3}$ pu and  zero sequence impedance of (0.031 + 0.103j) pu.
The t-line 4-9 is series compensated to improve its power transfer capabilities and other operational characteristics.
The faults are simulated at eight distinct locations by altering the priority mode, fault resistance, fault type, and fault inception angle. 
The generator at bus-3 is turned on and off to examine the effect of infeed in the case of capacitor and load-switching. Tables \ref{parameters1} and \ref{parameters2} provide the parameters and their values for simulating faults, load, and capacitor-switching scenarios. {A lg fault at location f1 with fault resistance 0.1 $\Omega$ and consisting of the dc decaying and ac components is shown in Fig. \ref{lg fault}.}

 \begin{figure}[ht]
 		\centering
	\captionsetup{justification=centering,textfont=small}
	\includegraphics[width=4.4 in, height= 2.7 in]{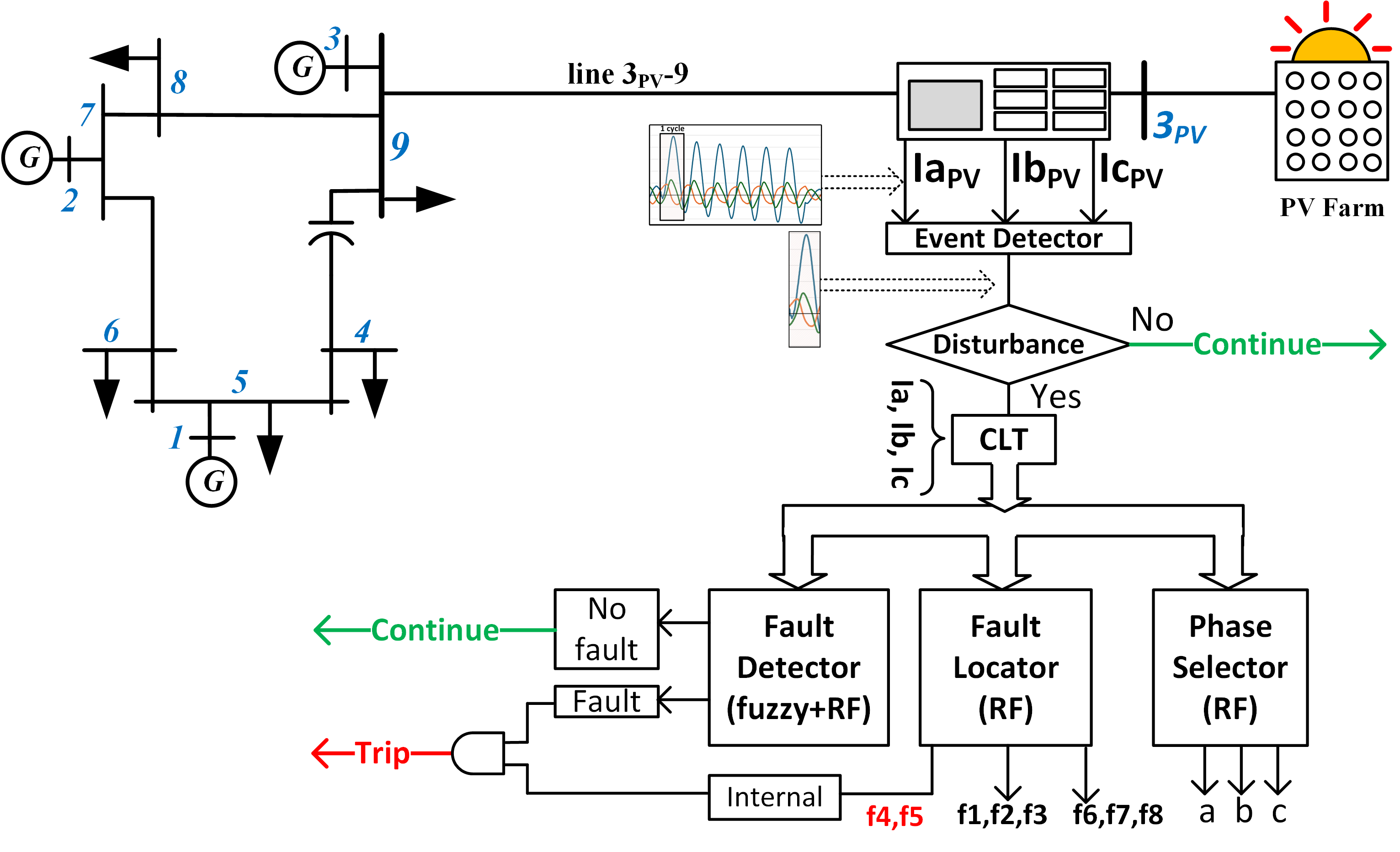}
	\caption{Framework for proposed hybrid protection system.}
	\label{fc}
\end{figure}

\section{Proposed Hybrid Protection System}
\subsection{Protection Scheme}
{Fig. \ref{fc}} provides an illustration of the suggested protection mechanism.  First, at the PV side of the t-line under consideration (line $3_{PV}$-9) the event detector checks for any anomalies in the 3-phase currents recorded by the $CT_{PV}$ and captures the data in case of transients. Linear trend attributes are extracted from the 1-cycle 3-phase currents that were recorded. 
Second, the CLT-trained fuzzy inference system and RF are used for fault detection. Third, line $3_{PV}$-9 is tripped if the fault locator identifies the captured transient currents as an internal fault (location f4 \& f5). Fourth, the faulty phase is detected. Preprocessing the data and extracting features are part of the first step. RF feature selection is used to rank the linear trend features.

\subsection{Event Detector}
An event detector (ED) is employed to identify any alterations in phase currents. It calculates the fractional increase by comparing the overall sum of the modulus of current samples from two consecutive current cycles (equation(\ref{eq_ed})).
\begin{equation}\
\label{eq_ed}ED (x) = \frac{\sum_{x=0}^{M}|I_{\phi}(x)|-\sum_{x=0}^{M}|I_{\phi}(x-M)|}{\sum_{x=0}^{M}|I_{\phi}(x)|} 
\end{equation}
where $M$ = no. of samples in a cycle, and $I_{\phi}$ = phase current.
The ED filter records the 3-phase current samples starting from the time instant $x$ that satisfies the equation (\ref{ed}).
\begin{equation}\label{ed}
 ED(x) \geq  \gamma = 0.06 \  \forall \  \phi \in a,b,c  
\end{equation}
The threshold $\gamma$ is determined by applying grey wolf optimization, a metaheuristic algorithm that explores for the optimal solution, drawing inspiration from the social hierarchy and hunting behaviors observed in packs of wolves \cite{greywolf}. Wolves are represented as candidate solutions, and their positions are updated iteratively based on fitness evaluations using an objective function. {The $\gamma$ value depends on the maximum fault resistance considered (here 10 $\Omega$). It decreases with an increase in fault resistance.}
Steps for identifying the optimal $\gamma$ involve:\\
{ I:  Initializing positions randomly with population size=25, dimension of search space=1, lower limit=0, upper limit=1, and   maximum iteration=200\\
 II: Assessing fitness values employing the objective function:\\
 \vspace{0.5mm}$(1 - \frac{ disturbances \ detected \ in \ 1 \ cycle }{ Total \ disturbances \ detected})$ \\
 III: Iterating and updating wolf positions according to dominance hierarchy. Assess the updated positions' fitness.\\
 IV: Tracking and returning the best solution and fitness value.}


\subsection{Random Forest Feature Selection}
Random Forest is used to select and rank the features in which
multiple decision trees are trained on a random subset of the data. A random subset of the features at each node of the decision tree is chosen for splitting.
 For each feature at a node, the Gini impurity is calculated before and after the split. The difference in impurity is used to assess the importance of that feature. The feature importance scores are averaged over all decision trees. Finally, features are ranked on the average Gini impurity reduction, with higher values indicating more important features.
\begin{equation} Gini(n) = 1 - \sum_{i=1}^{C} (p(i|n))^2
\end{equation}
$Gini(n)$ is Gini impurity at node $n$, $C$ is number of classes, $p(i|n)$ is probability of class $i$ at node n.
RF feature selection chooses the most significant linear trend feature (Table \ref{featurelist}).

\subsection{Linear Trends}
The 3-phase relay currents can be characterized using pertinent extracted features, which unveil unconventional insights into the dynamics of the fault currents \cite{pedes}. Linear trends are used in various real-world applications across different fields like finance, climate science, healthcare, engineering, etc. to analyze and predict patterns and relationships. Linear trends were used to distinguish faults and transients in a 5-bus power network in \cite{systempallav} and to detect faults during power swings and classify
power swings in \cite{powerswing}.

\textit{Simple linear trend (SLT):}
A linear trend, in the context of time series analysis, refers to a consistent and steady increase or decrease in the values of a variable over time. It represents a straight line pattern when plotted on a graph, where the data points follow a linear relationship as time progresses.  For the time series values, it computes the linear least-squares regression, and “p-value”, “correlation coefficient”, “intercept”, “slope”, and “standard error” are obtained \cite{tsfresh}. In this scenario, the three time series correspond to the 3-phase currents monitored by $CT_{PV}$.

The linear trend equation is given by:
\begin{equation} f = mg + b \end{equation}
where y-intercept is \(b\), slope of the straight line is \(m\), dependent variable is \(f\), and independent variable is \(g\).

To find the best-fitting values of \(m\) and \(b\), ``least squares regression" is used. The purpose is minimizing the sum of squares:
\begin{equation} S = \sum_{i=1}^{n} (f_i - (mg_i + b))^2 \end{equation}
where \((g_i, f_i)\) are the data points for \(i = 1, 2, ..., n\).

To find the optimal values of \(m\) and \(b\), the partial derivatives of \(S\) with respect to \(m\) and \(b\) is taken and set to zero:
\begin{equation} \frac{\partial S}{\partial m} = 0 \end{equation}
\begin{equation} \frac{\partial S}{\partial b} = 0 \end{equation}

Solving these equations simultaneously gives the values of \(m\) and \(b\) minimizing the sum of squares:

\begin{equation} m = \frac{n\sum (f_i h_i) - \sum g_i \sum f_i}{n\sum (f_i^2) - (\sum g_i)^2} \end{equation}

\begin{equation} b = \frac{\sum f_i - m\sum g_i}{n} \end{equation}

Once these values are obtained the linear equation \(f = mg + b\) is used to predict the values of \(f\) for any given \(g\). The linear trend line will represent the ``best fit" line through the data points, minimizing the overall distance between the actual values and the expected values.

Pearson's correlation coefficient (Pearson's \(r\)) can then be obtained for the data points using:
\begin{equation}
r = \frac{\sum (g_i - \overline{g})(f_i - \overline{f})}{\sqrt{\sum (g_i - \overline{g})^2 \sum (f_i - \overline{f})^2}}
\end{equation}
where \(g_i\) and \(f_i\) are the individual data points 
, \(\overline{g}\) and \(\overline{f}\) are the means of the variables \(g\) and \(f\), respectively. The summation is over all data points in the dataset.

\begin{table}[ht]
	\footnotesize
	\renewcommand{\arraystretch}{1.0}
	\setlength{\tabcolsep}{4 pt}
	\centering
	\captionsetup{justification=centering}
	\caption{{List of linear trend features extracted} }\label{featurelist}
	\label{featuretable}
	\begin{tabular}{|c|c|c|}
		\hline
		\rowcolor[rgb]{0.91,0.91,0.91} $Feature$       & $Description$                                       &     $Parameters$                                                                              \\ \hline
		\begin{tabular}[c]{@{}c@{}}Simple\\Linear Trend\\(1-5)\end{tabular}       & \begin{tabular}[c]{@{}c@{}}calculates least-squares \\ regression for time series values\end{tabular}  & \begin{tabular}[c]{@{}c@{}}`p-value', `Pearson's r', \\ `intercept', `slope', `error'\\total = 5\end{tabular} \\ \hline
		\begin{tabular}[c]{@{}c@{}}Combined \\Linear Trend\\(6-65) \end{tabular} & \begin{tabular}[c]{@{}c@{}} calculates least-squares \\ regression for time series \\values combined over segments  \end{tabular}                                             & \begin{tabular}[c]{@{}c@{}}`p-value', `Pearson's r', \\ `intercept', `slope', `error'\\\  Segment size = 5, 10, 50\\ $f$=mean, variance, max, min \\total = $ 5 \times 3 \times 4=60$\end{tabular}                             \\ \hline
	\end{tabular}
\end{table}

\begin{figure}[htbp]
	\centering
	\begin{minipage}[b]{0.40\textwidth}
		\centering
		\includegraphics[width=\textwidth]{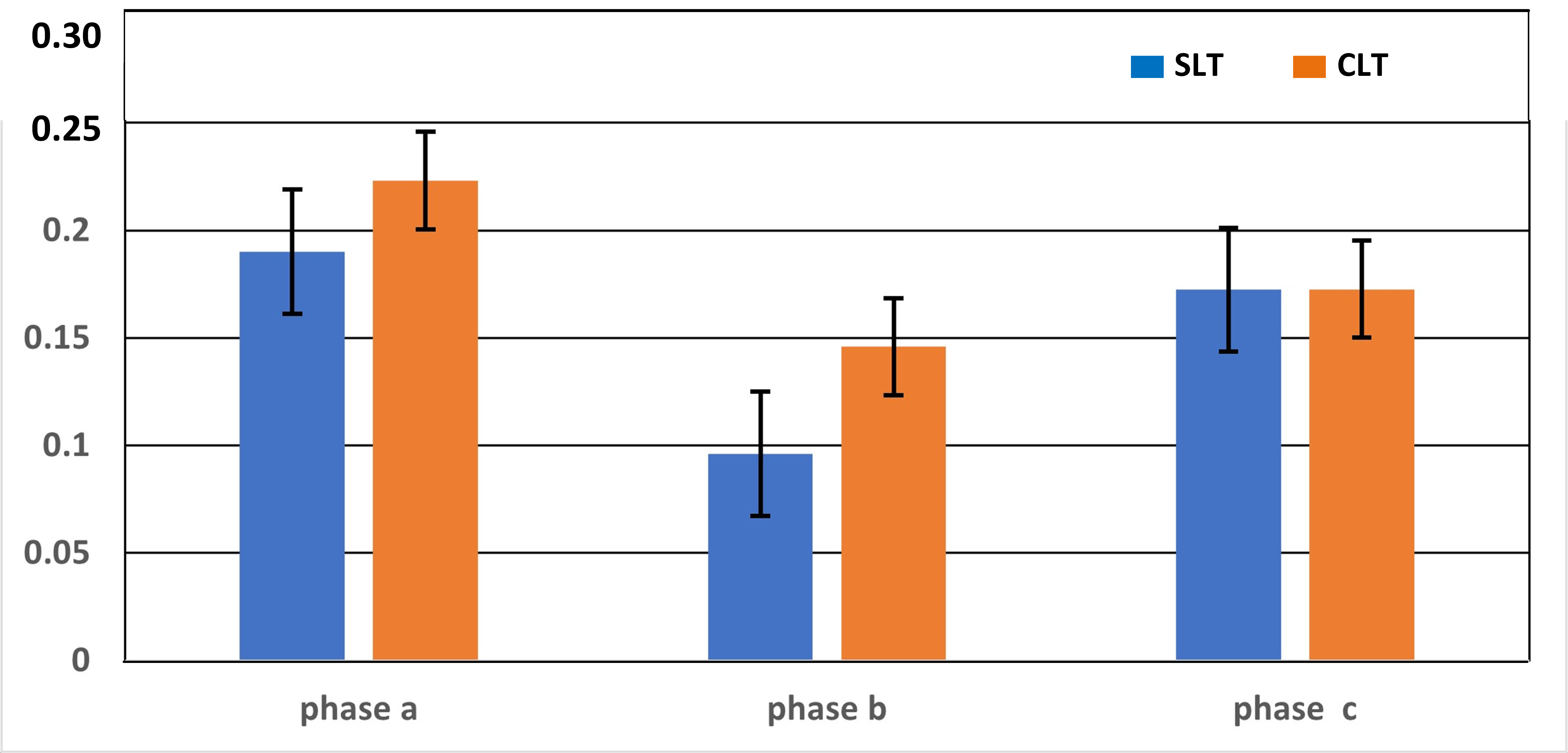}
		\captionsetup{font=footnotesize}
		\caption{Feature importance of SLT and CLT}
		\label{feature importance}
	\end{minipage}
	\hfill
	\begin{minipage}[b]{0.54\textwidth}
		\centering
		\includegraphics[width=\textwidth]{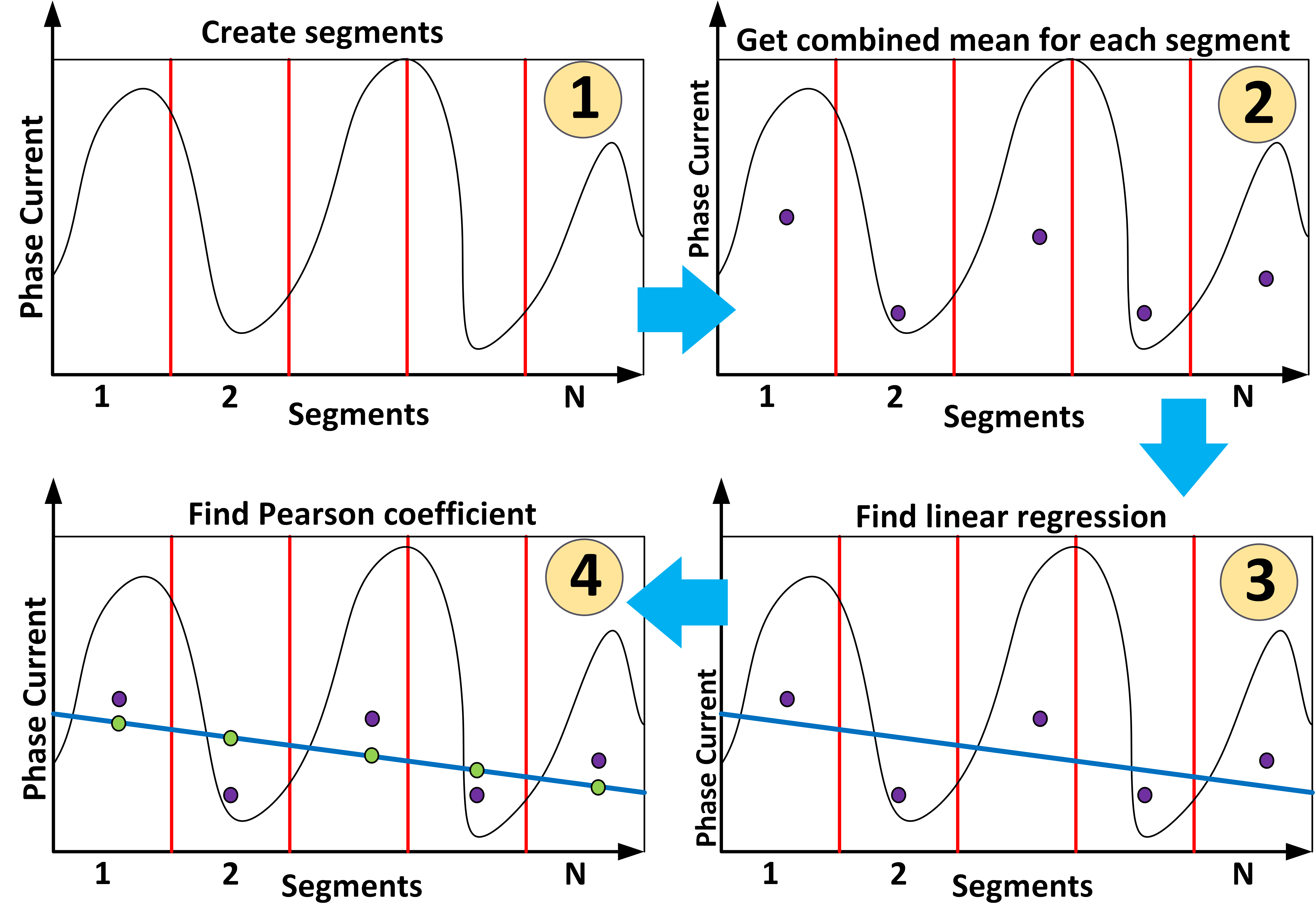}
		\captionsetup{font=footnotesize}
		\caption{Steps for calculation of combined linear trend (CLT)}
		\label{clt}
	\end{minipage}
\end{figure}

\textit{Combined linear trend (CLT):}
In addition to SLT attributes, the CLT attributes are also evaluated. For time series values aggregated across segments, CLT performs a linear least-squares regression. 
 The same attributes:  “p-value”, “correlation coefficient”, “intercept”, “slope”, and  “standard error” are extracted with the segment size varied between 5, 10, and 50. The number of time series values in each segment is determined by the segment size.   For a segment size of 10, the number of segments is 12.8 $\approx$ 13 (1 cycle = 128 data points), thus reducing the data points from 128 to 13. Maximum, minimum, mean, or variance of time series values in a segment is used to get the combined value.  

 The list of SLT and CLT features evaluated using RF feature selection is listed in Table \ref{featurelist}. CLTs with attribute: `Pearson's r', segment size: 50, and $f$: mean for the 3-phase currents are thus chosen after feature selection. The importance of SLT and CLT are shown in Fig. \ref{feature importance}.  In determining the feature importance of (say $SLT\ phase\ a$) from a tree, the process involves first computing the importance specifically for nodes where the split occurred due to feature $ SLT\ phase\ a$, then dividing it by the total feature importance of all nodes. The RF feature importance is derived by averaging the importances across all trees. It is apparent from the figure that the CLT is a more effective feature for identifying faults in t-lines connecting PV farms.


The feature calculation steps for CLT are displayed in Fig. \ref{clt}. First, the one-cycle phase currents are split into N = 50 segments. Second, the data of each segment is aggregated over the mean in a single data point per segment to reduce the number of measurement points to the number of segments.
Third, the acquired data points are then used to create a linear least-squares regression line. Fourth, the Pearson correlation coefficient is utilized to describe the data points. Here, the feature is named correlation coefficient $r$ which is dimensionless.


\subsection{Random Forest}

Once the input features (CLT) are selected, they are used to train Decision Tree (DT), Random forest (RF), Naive Bayes (NB), Support-vector machines (SVM), and k-nearest neighbors (kNN) classifiers. 

RF is a classification algorithm that uses ensemble learning. It creates multiple decision trees through bootstrapping (random subsets of data) and random feature selection. Every tree makes a class label prediction; the ultimate prediction is decided by a majority vote of all the trees. This approach improves accuracy and reduces overfitting, making RF a powerful and widely used classifier.
The predicted class for a given input sample \( p\) is depicted as:

\begin{equation}
   \hat{q} = \text{argmax}_i \sum_{j=1}^{N_T} I\left( T_j(p) = i \right) 
\end{equation} 
where \( \hat{q} \) is predicted class label for the input sample \( p \), \( N_T \) is total number of decision trees, 
\( T_j(p) \) is predicted class label by the \( j \)-th decision tree.
The indicator function \( I \) returns 1 in the case that the condition included in parenthesis is true and 0 in the other case.

  \begin{minipage}[b]{0.51\textwidth}
	\vspace{5mm}
	\centering
	\includegraphics[width=\textwidth]{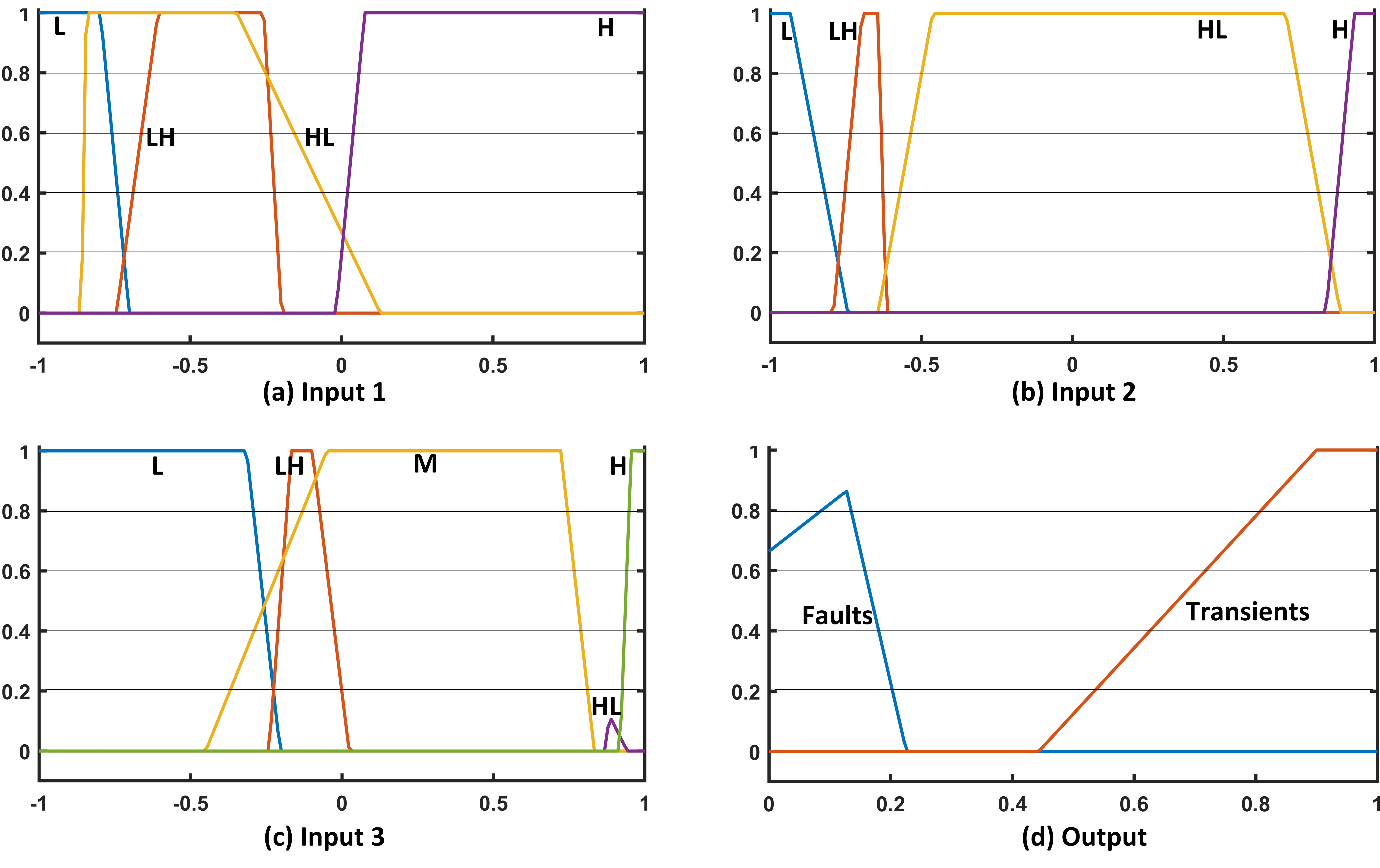}
	\captionsetup{font=footnotesize} 
	\captionof{figure}{GA tuned Mamdani fuzzy system having 3 inputs and 1 output with trapezoidal membership functions for fault detection}
	\label{fuzzy} 
\end{minipage}  
\hspace{0.02\textwidth}
\begin{minipage}[b]{0.44\textwidth}
	\centering
	\scriptsize
	\renewcommand{\arraystretch}{1.2}
	\setlength{\tabcolsep}{1 pt}
	\captionsetup{font=footnotesize} 
	\captionof{table}{Fault Detection performance with chosen five traditional ML algorithms}
	\label{fd}
	\begin{tabular}[b]{|l|l|l|l|l|l|} \hline
		\backslashbox{\textit{Accuracy($\bar\eta$)}}{\textit{Classifier}} & \textit{SVM} & \textit{RF} & \textit{DT} & \textit{kNN} & \textit{NB} \\ \hline                                                  
		\textit{without} SMOTE (\%) & 93.3   & \textbf{98.0}  & 96.2  & 96.5   & 70.3 \\ \hline
		\textit{with} SMOTE (\%)    & 92.6   & \textbf{98.2} & 97.0  & 97.0   & 70.1 \\ \hline
	\end{tabular}  
\end{minipage}

\section{Findings of Detection, Location, and Phase Selection of Faults}
This section examines the outcomes of fault detection, fault location, and phase selection stages on the IEEE 9-bus system.
 For evaluation of ML algorithms the fault and transient dataset is divided into two separate subsets: training set and test set in a 4:1 split ratio to strike a balance between training and evaluation, ensuring models do not overfit and are proficient in handling new, unseen data. 
 Stratified 10-fold cross-validation is used which combines stratification and cross-validation, providing a robust means to thoroughly assess model performance. Stratification groups data by class labels to maintain class distribution and 10-Fold Cross-Validation divides data into 10 subsets. Iterates 10 times, training on 9 subsets and testing on 1, then averages performance for reliable metrics. Grid search is applied for hyperparameter optimization 
 which explores exhaustively all possible combinations of the hyperparameter values in a specified range to determine the combination that yields the best performance.
The common metric used to assess the classifiers' performance is accuracy.  However, it is skewed toward data imbalance. Hence, balanced accuracy $\bar{\eta}$ is used to gauge the model's performance where $\bar{\eta}$ = $ \frac{1}{2}[\frac{TP}{(TP+FN)} + \frac{TN}{(TN+ FP)}]$ for a binary class problem in which the terms true positive (TP), false negative (FN), false positive (FP), and true negative (TN) are used.
Further, SMOTE (synthetic minority over-sampling technique) \cite{smote}, which aims to alleviate the issue of unbalanced datasets by generating synthetic samples of the minority class, is also applied. 



\begin{figure}[htbp]
	\centering
	\begin{minipage}[b]{0.53\textwidth}
		\centering
		\includegraphics[width=\textwidth]{3dplot2.pdf}
		\captionsetup{font=footnotesize}
		\caption{Surface plots showing hyperparameter search for RF}
		\label{sp}
	\end{minipage}
	\hfill
	\begin{minipage}[b]{0.44\textwidth}
		\centering
		\includegraphics[width=\textwidth]{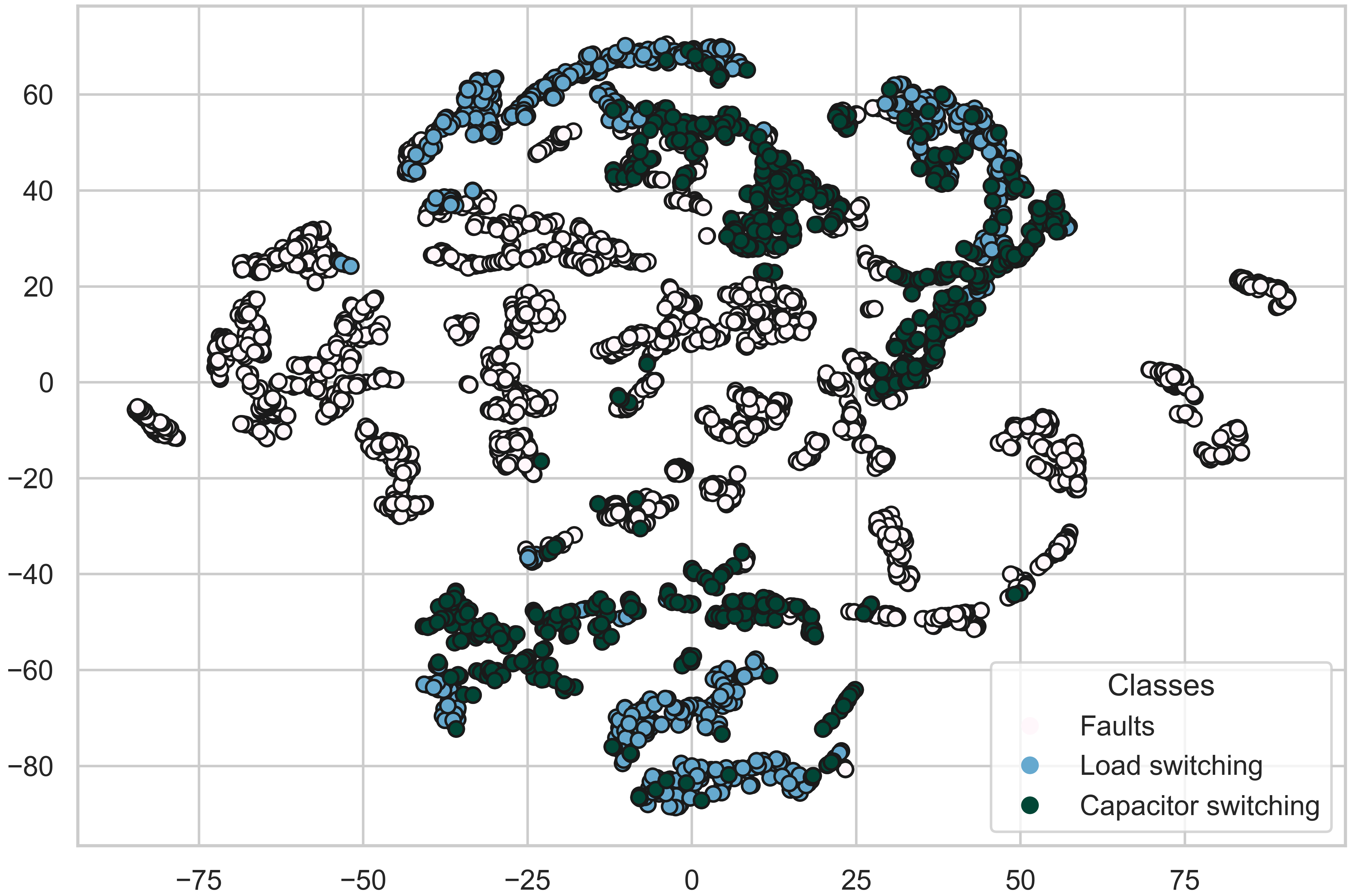}
		\captionsetup{font=footnotesize}
		\caption{Scatter plot showing CLTs of faults, load-switching, and capacitor-switching cases}
		\label{tsnefig}
	\end{minipage}
\end{figure}

  \begin{figure}[ht]
	\centering
	\includegraphics [width=3.5 in, height= 2.55 in] {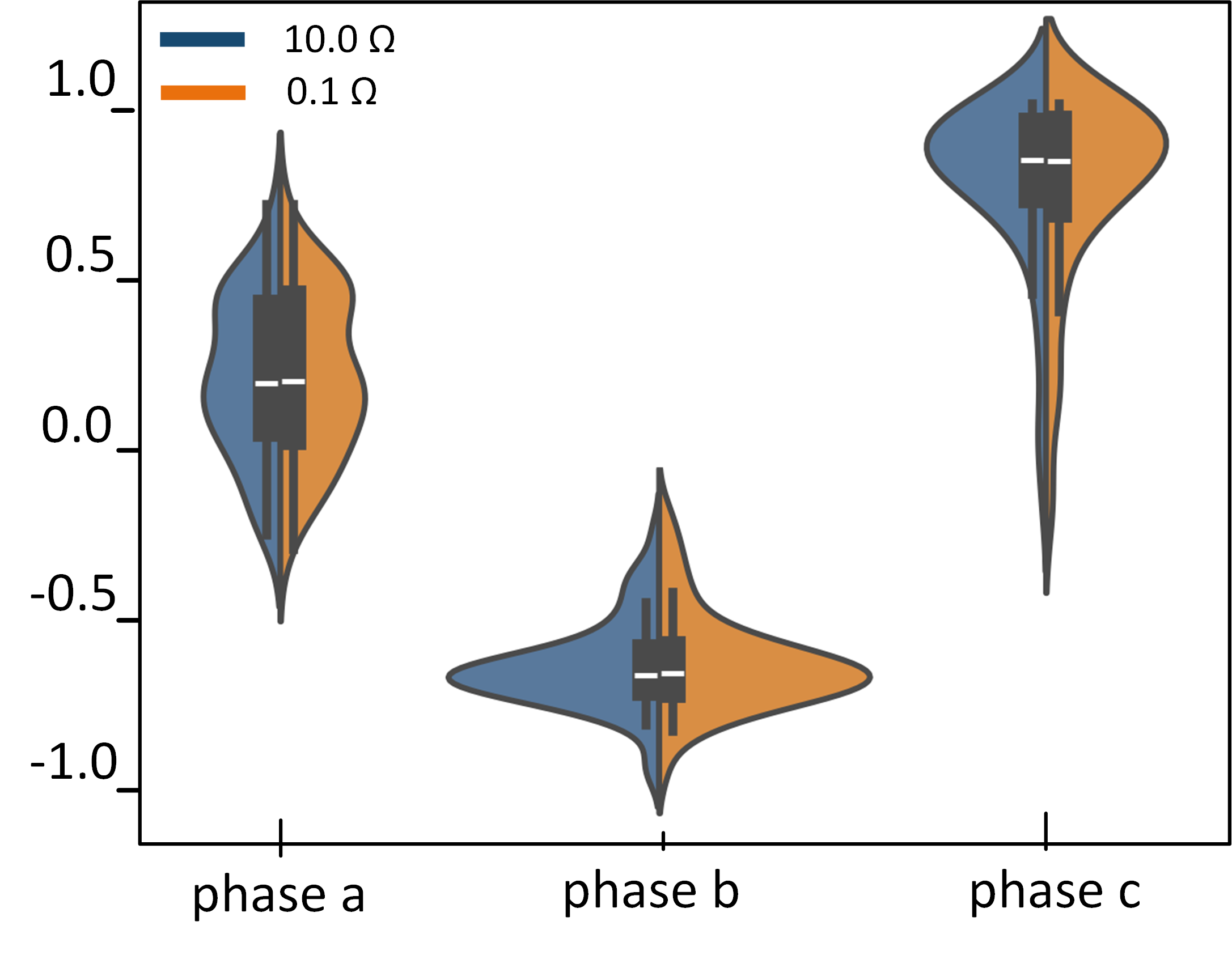} 
	\caption{Spread of CLT features for 0.1 $\Omega$ and 10 $\Omega$ fault resistances at location f4 on 60 fault cases each considering different fault types and fault times}\label{fault resistances}
\end{figure}

\subsection{Fault Detection}
The fault detector is a hybrid intelligent system consisting of a fuzzy system supported by ML. The fuzzy logic system provides a framework for handling uncertainty and imprecision in the decision-making process. {The a,b,c,d parameters of the trapezoidal membership function define it's shape  and the fuzzy rules determine how the fuzzy sets interact to produce outputs.} Genetic Algorithm is used to fine-tune these parameters of the trapezoidal membership functions of inputs (CLTs of 3-phase currents) and output for optimal performance (Fig. \ref{fuzzy}). The fuzzy inference system customized using a data-driven methodology is used to find the faults.


In research articles on power system protection, RF, SVM, DT, kNN, and NB have shown promising outcomes. The classifiers take as inputs the 3-phase CLTs that were chosen using RF feature selection. Table \ref{fd} shows the $\bar\eta$ for different classifiers with SMOTE and without SMOTE analysis for fault detection. RF outperforms the SVM, DT, kNN, and NB. The hyperparameters for these classifiers are optimized using grid search.
It is observed that the use of SMOTE didn't influence the results considerably. RF classifier gives the best $\bar\eta$ of 98.0\% without SMOTE (2880 faults and 2400 non-faults) and $\bar\eta$ of 98.2\% with SMOTE (2880 faults and 2880 non-faults). The optimal hyperparameters of RF are obtained with grid search on n\_estimators = [400, 800, 1200, 1600, 2000, 2400, 2800, 3200, 3600, 4000], min\_splits = [2, 3, 4, 5, 6, 7, 8, 9, 10], and  max\_depth =[2, 4, 6, 8, 10, 12, 14, 16]. The best hyperparameter (n\_estimators = 400, min\_splits = 2, max\_depth = 14) obtained is depicted in the 3D surface plot in Fig. \ref{sp}. The 3D surface plots help understand the relationship between any two out of the three RF hyperparameters and model performance in the hyperparameter optimization.
The higher accuracy of RF compared to other algorithms can be attributed to ensemble learning, feature importance, robustness to overfitting, and ability to handle high-dimensional and non-linear data.



t-SNE \cite{tsne} plot is used to visualize the high-dimensional data in a lower-dimensional space while preserving the local structure and relationships between data points (correlation coefficient of CLTs) of faults, load-switching, and capacitor-switching events (Fig. \ref{tsnefig}). It reveals clusters of capacitor and load-switching transients with scattered clusters of fault data having higher variability. The separation of fault and switching transient clusters visible in the plot makes it easier for the ML classifiers to differentiate them.

{It's essential that the suggested algorithm works well for various fault resistances. Box plots offer a clear and detailed depiction of a dataset’s distribution encapsulating key data points – the minimum, first quartile, median, third quartile, and maximum – providing a comprehensive view of the data’s spread, central tendency, skewness, and potential outliers. The side by side comparisons of each feature for 0.1$\Omega$ and 10$\Omega$ using Violin plots which provide the shape of the distribution in addition to the information from boxplots are shown in Fig. \ref{fault resistances}. It is evident that the faults with 0.1$\Omega$ resistance have slightly more variability or spread (width of IQR and length of whiskers) than 10$\Omega$. However, the density plots are similar. This overall similarity in density and box plots for all three features suggests that a classifier would be effective in distinguishing fault cases with different resistances from switching transients.}




 \begin{minipage}[b]{0.44\textwidth}
	\vspace{5mm}
	\centering
	\scriptsize
	\renewcommand{\arraystretch}{1.2}
	\setlength{\tabcolsep}{1 pt}
	\captionsetup{font=footnotesize} 
	\captionof{table}{Results for Fault Location and Phase Selection}
	\label{loc}
	\begin{tabular}[b]{|l|l|l|l|l|l|} \hline
		\backslashbox{\textit{Accuracy($\bar\eta$)}}{\textit{Classifier}} & \textit{SVM} & \textit{RF} & \textit{DT} & \textit{kNN} & \textit{NB} \\ \hline   
		\textit {Fault location (\%)}
		& 80.5            & \textbf{92.3}          & 88.3           & 89.2            & 69.5           \\ \hline
		\textit {Phase selection (\%) } & 97.0            & \textbf{97.2}           & 85.3           & 88.5            & 68.4           \\ \hline
	\end{tabular}
\end{minipage}
\hspace{0.03\textwidth}
\begin{minipage}[b]{0.48\textwidth}
	\centering
	\includegraphics[width=\textwidth]{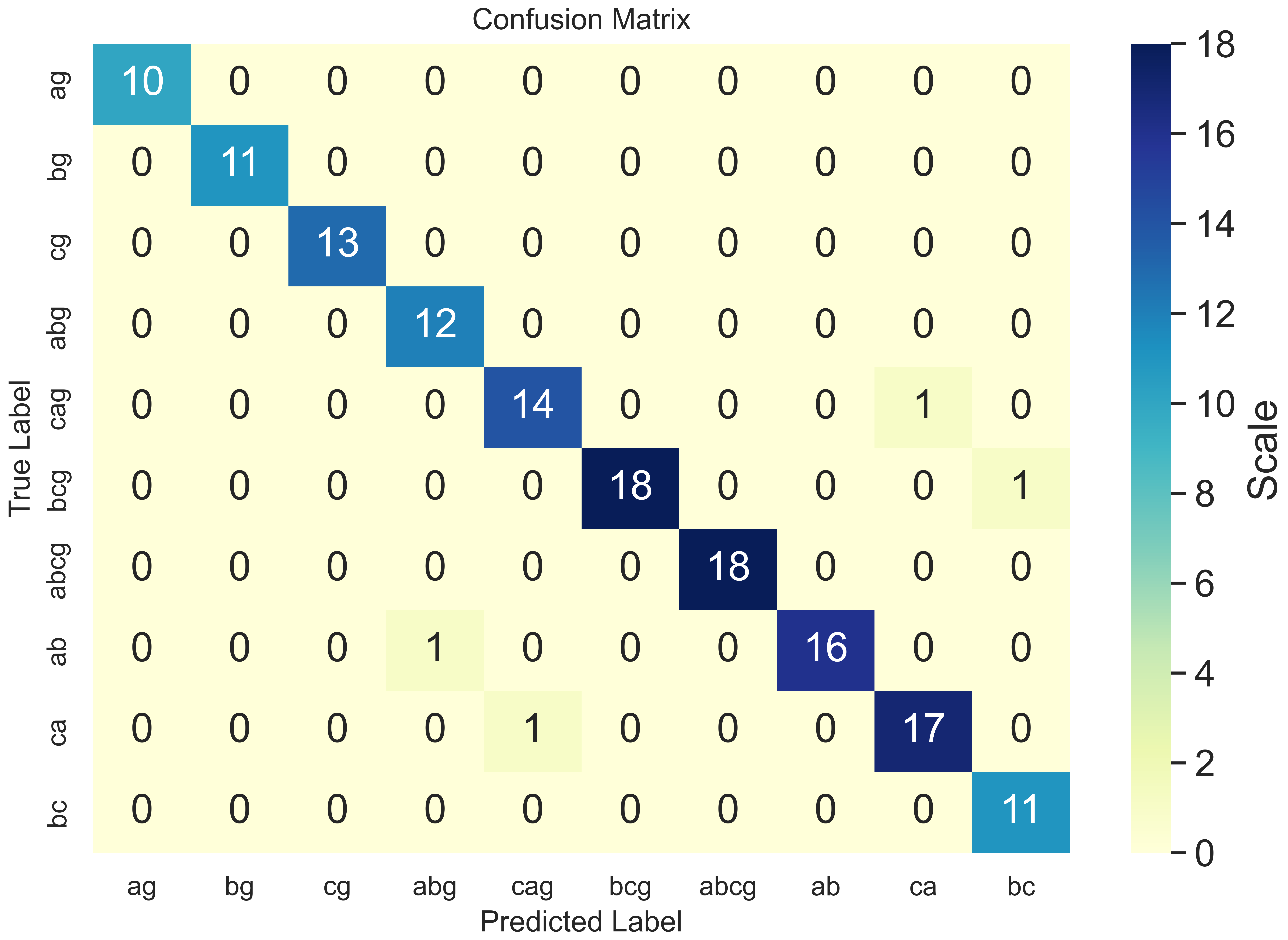}
	\captionsetup{font=footnotesize} 
	\captionof{figure}{Confusion/Error Matrix for phase selection}
	\label{cf} 
\end{minipage}

\subsection{Fault location}
Once a transient is identified as a fault, the proposed method ascertains the  region of fault. Eight places are used to simulate these faults, with internal faults at f4 and f5. Faults at locations f1, f2, f3 are considered as backward external faults, and f6, f7, and f8 as forward external faults. Table \ref{loc} shows the $\bar\eta$ for different classifiers with RF outperforming the others with a $\bar\eta$ of 92.3\% on 2880 (see Table \ref{parameters1}) fault cases.


\subsection{Phase selection}
Post fault detection and identification of the fault as internal, the faulty phase is determined. The faulty phase/phases among phase a, phase b, phase c, phases ab, phases bc, phases ca, or phases abc are identified. Again, RF gives $\bar\eta$ of 97.2\% on 720 faults at f4 and f5 locations (Table \ref{loc}). The phase selection confusion/error matrix is displayed in Fig. \ref{cf}.




\section{Validity of Proposed Scheme in Different Scenarios}
The next paragraphs discuss the probable scenarios that could create challenges for the suggested CLT-based approach. The capability of RF to detect faults and switching transients is assessed for these unique circumstances.

\begin{figure}[ht]
	\centering
	\includegraphics [width=3.6 in, height= 3.4 in] {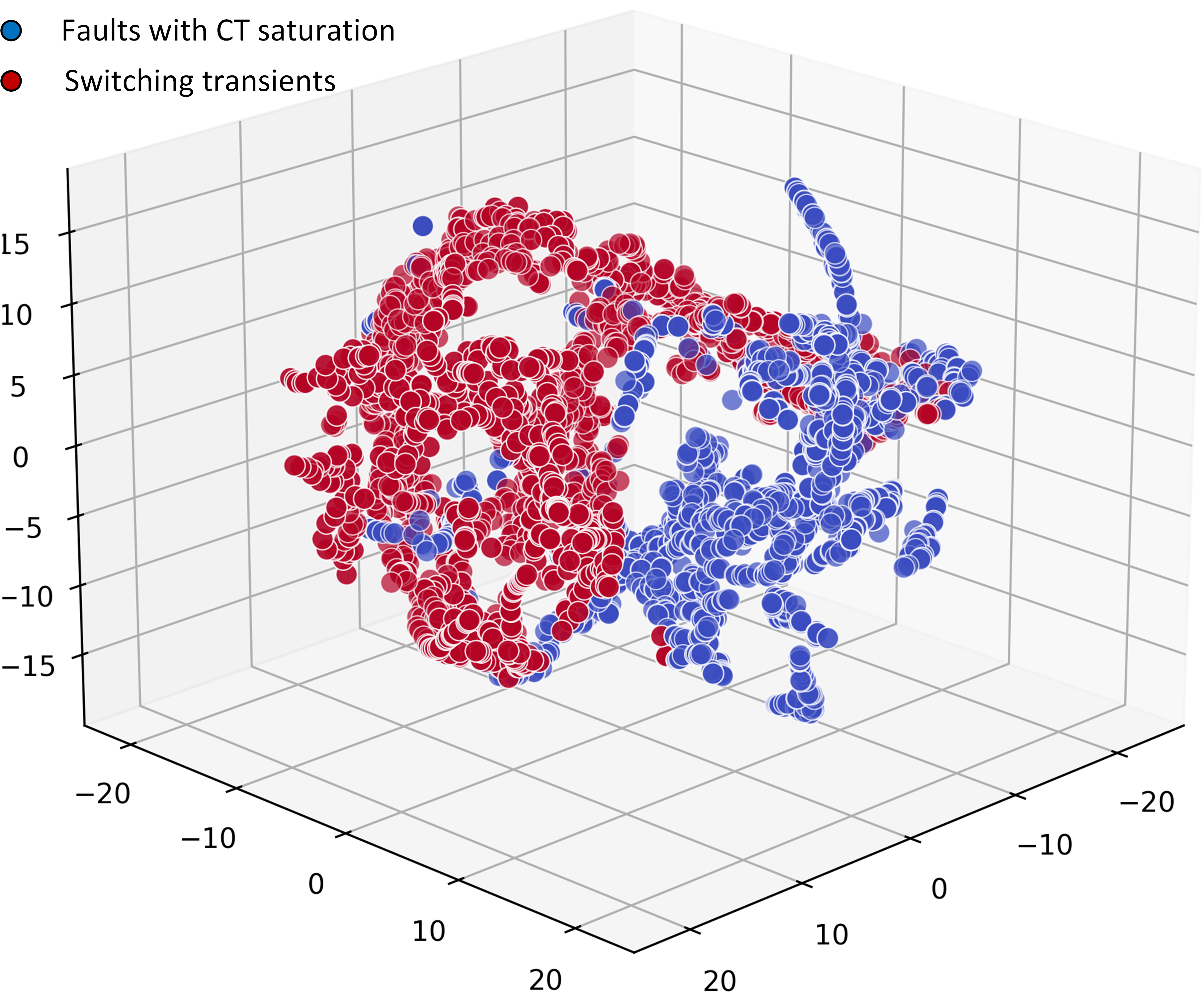} 
	\caption{3D tSNE plot showing the CLT coefficients for faults with CT saturation and switching transients}\label{ctsat3d}
\end{figure}
\begin{figure}[ht]
	\centering
	\includegraphics [width=4.5 in, height= 3.6 in] {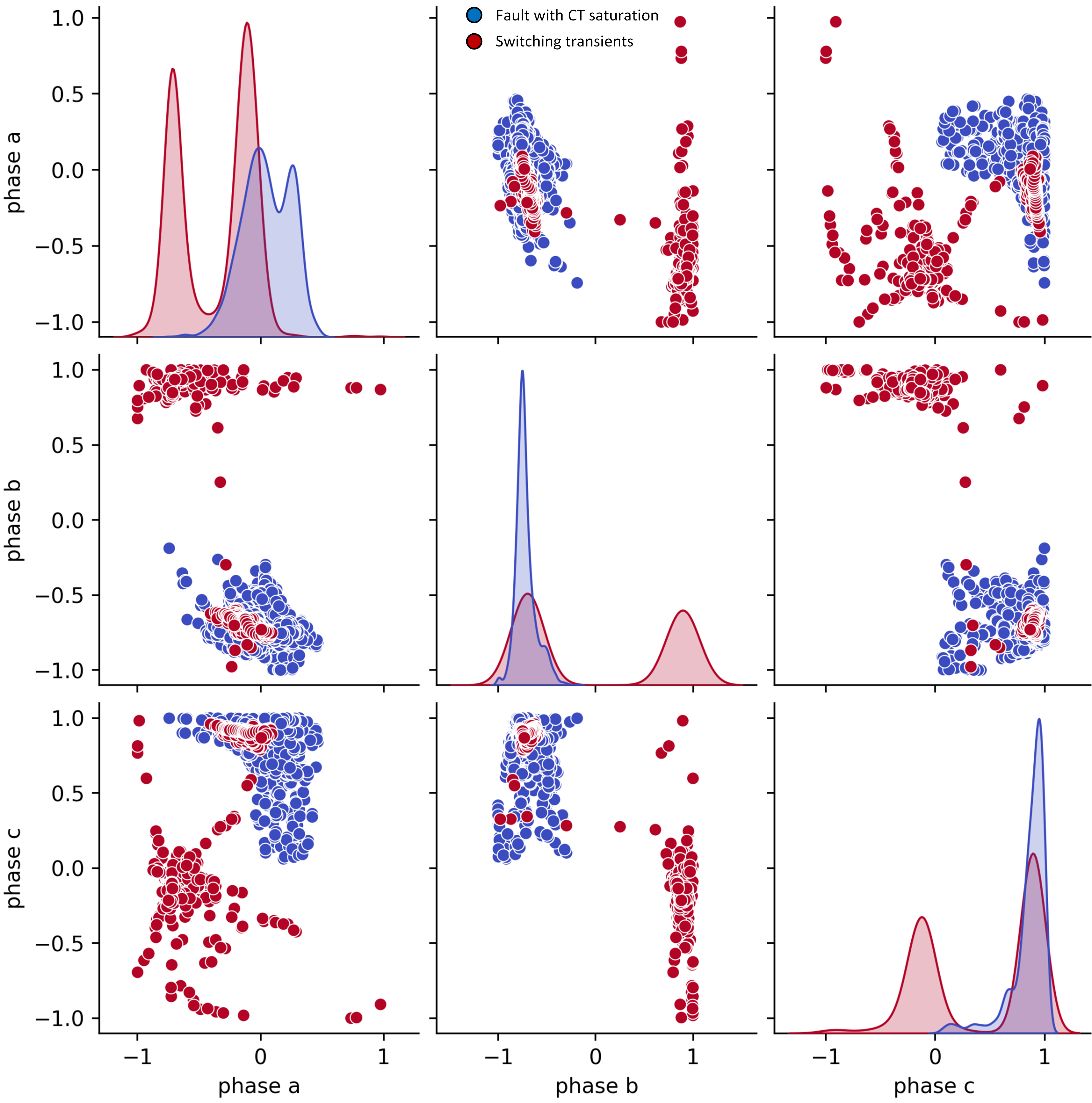} 
	\caption{Pair plot for the 3-phase CLT features}\label{pair plot}
\end{figure}

\subsection{\textbf{Effect of Faults with CT Saturation and CVT Transients}}
During severe faults, the CT cores may get saturated, which could negatively impact how well protection algorithms perform. A fault with CT saturation has no effect on the CLT-based fault detection.
To show this the secondary burden is increased to 20 ohms, which will cause the CTs to become saturated. The faults are identified with $\bar\eta$ of 97.3\% on 1760 faults simulated in P and Q priority for fault resistances (0.01, 1, 2, \& 10 ohms), fault types (10), and fault inception angles (11) at fault locations (f5 \& f6); and 2400 switching transients cases. {For the fault and switching transients, representing two distinct classes, the 3D t-SNE plot illustrates how effectively the classes are separated in the reduced space, showcasing the feature's discriminative power. The clusters do not overlap, indicating that the selected features can distinguish between the classes (Fig.\ref{ctsat3d}). Additionally, in Fig.\ref{pair plot}, the diagonal elements display the distribution of each individual feature. For example, the cell at row phase a and column phase a shows the density plot of the phase a CLT feature. The off-diagonals present pairwise scatter plots of the features. For instance, the cell at phase a row and phase b column is a scatter plot comparing phase a with phase b CLT feature.}
CVT transients, which typically cause overreach for zone 1 distance relays with no intentional delay \cite{cvt}, have no effect on the proposed method because only phase currents are taken into account here for feature extraction.

\subsection{\textbf{Effect of Change in Farm Capacity}}
By changing the PV units from 400 (base value) to 300 and 500, the PV system's capacity is altered, and the suggested method is evaluated.
The technique recognized the faults and switching transients with $\bar\eta$ of 96.0\% and 96.4\% on 300 and 500 units respectively using 2400 no-fault transients and 2160 faults simulated by changing priority (P and Q), fault resistances (0.01, 1, and 10 ohms), fault types (10), and  fault inception angles (6) at locations f2, f5, and f6. 

\begin{table}[htbp]
	\footnotesize
	\centering
	\begin{minipage}[b]{0.45\textwidth}
		\centering
		\captionsetup{font=small} 
		\caption{Effect of noise in the 3-phase currents}
		\label{noise}
		\begin{tabular}{|l|c|}
			\hline
			\textbf{Noise(dB)} & \textbf{Accuracy(\%)} \\ \hline
			$\infty$        & 98.0     \\ \hline
			40        & 96.3       \\ \hline
			30        & 95.8       \\ \hline
			20        & 94.3       \\ \hline
		\end{tabular}
		
	\end{minipage}
	\hfill
	\begin{minipage}[b]{0.45\textwidth}
		\centering
		\captionsetup{font=small} 
		\caption{Effect of sampling rate of the 3-phase currents}
		\label{fs}
		\begin{tabular}{|l|c|}
			\hline
			\textbf{Sampling(kHz)} & \textbf{Accuracy(\%)} \\ \hline
			7.68         & 98.0     \\ \hline
			5.76 (3/4)       & 96.5       \\ \hline
			5.12  (2/3)     & 74.1       \\ \hline
			3.84 (1/2)       & 70.2       \\ \hline
		\end{tabular}
		
	\end{minipage}
\end{table}

\begin{figure}[ht]
	\centering
	\includegraphics [width=5.0 in, height= 2.2 in] {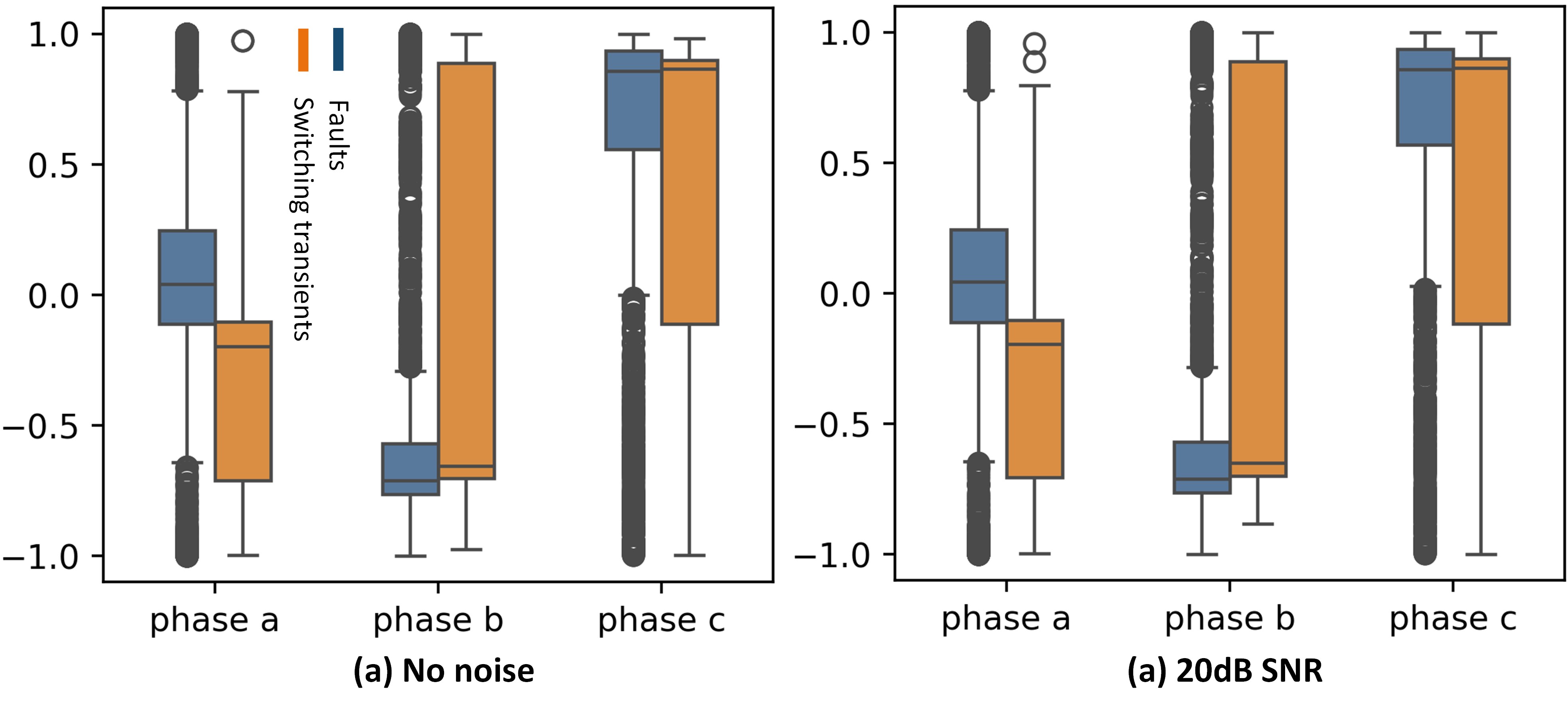} 
	\caption{Spread of CLT features for different noise levels}\label{noises}
\end{figure}

\subsection{\textbf{Performance in Presence of Noise}}
Measurement noise can significantly impact the performance of protection methods in power systems by leading to false tripping, missed fault detection, coordination issues, and more.
Electromagnetic interference from nearby equipment, power lines, and other sources can cause noise in current waveforms.
A 220 kV t-line normally has a 25 dB noise level. CLTs from 3-phase currents with SNRs of 20 dB, 30 dB, and 40 dB are used to test the anti-noise capability of the suggested method. The 3-phase currents obtained at $CT_{PV}$ have Gaussian white noise added to them. 
The $\bar\eta$ decreases from 98\% for no noise to 94.3\% for 20 dB noise (Table \ref{noise}).
{CLT features for 2880 fault and 2400 switching transients without noise and with 20dB noise show that switching transient have more spread and skewness (unequal whiskers or median line closer to Q1 or Q2) than the fault data. However, the boxplots at 20db are very similar to the ones for no noise (Fig. \ref{noises}).}

\subsection{\textbf{Effect of Sampling Frequency and Window Length}}
The relay's ability to operate quickly and reliably can be impacted by the frequency of data sampling and length of the data window. 
High sampling frequencies capture more detailed information about the power system's behavior, including fast transients and high-frequency disturbances.
Low sampling frequencies may miss critical events and nuances in the system's behavior, potentially leading to false alarms or delayed responses. However, a high sampling rate generates larger volumes of data, which can be challenging to store, process, and transmit in real time.
Therefore, a trade-off between data processing and accuracy is necessary.
Multiple kilohertz (kHz) are used to sample the 3-phase relay currents in order to assess the effects of sampling frequency. 
The quantity of samples utilized to train the RF classifier influences the suggested method. It is observed that the CLT-based scheme suffers when the sampling rate is reduced below 5.76 kHz (Table \ref{fs}). 

The optimal data window size depends on the characteristics of transients and the desired trade-offs between temporal resolution and computational efficiency.
The impact of the data window is examined using window sizes of half, one, and two cycles. $\bar\eta$s of 98.0\%, 98.0\%, and 97.2\%  are obtained correspondingly, showing the scheme's resilience to change in window size.

 \begin{figure}[ht]
	\centering
	\includegraphics [width=3.8 in, height= 2.5 in] {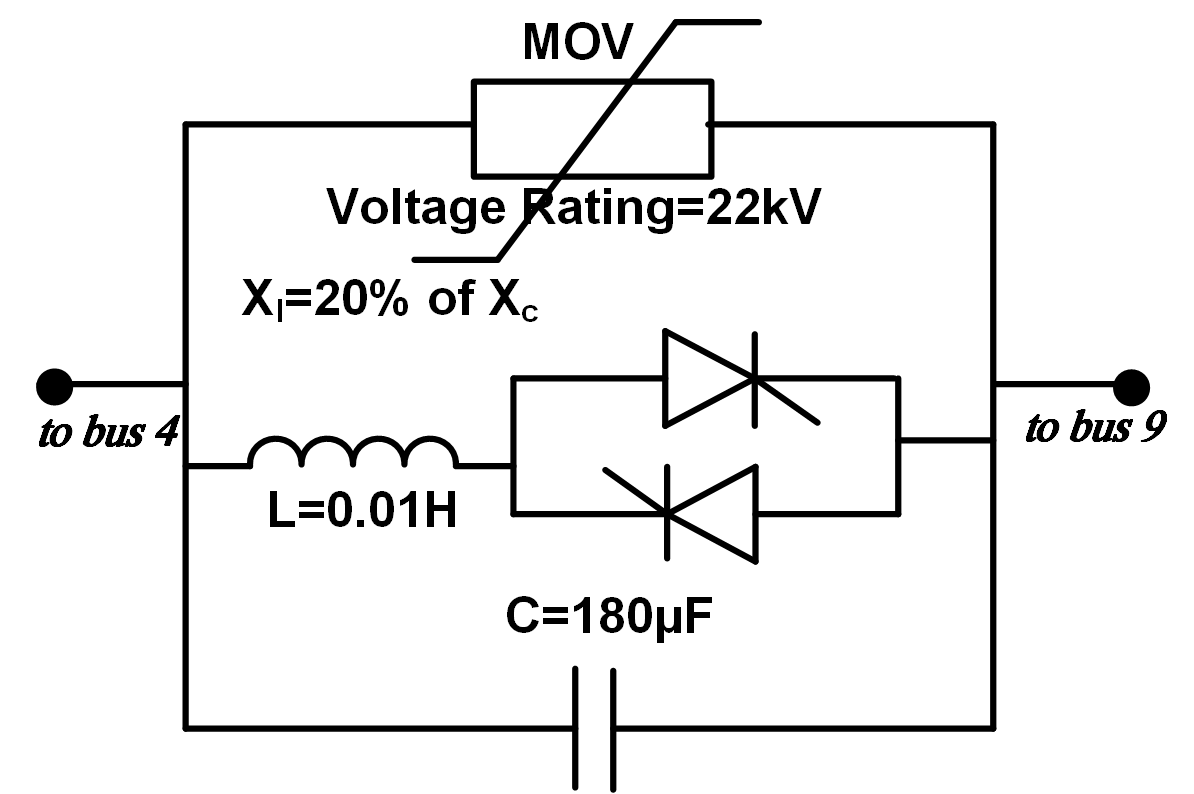} 
	\caption{Configuration of TCSC}\label{tcscfig}
\end{figure}

\begin{figure}[ht]
	\centering
	\includegraphics [width=3.2 in, height= 2.4 in] {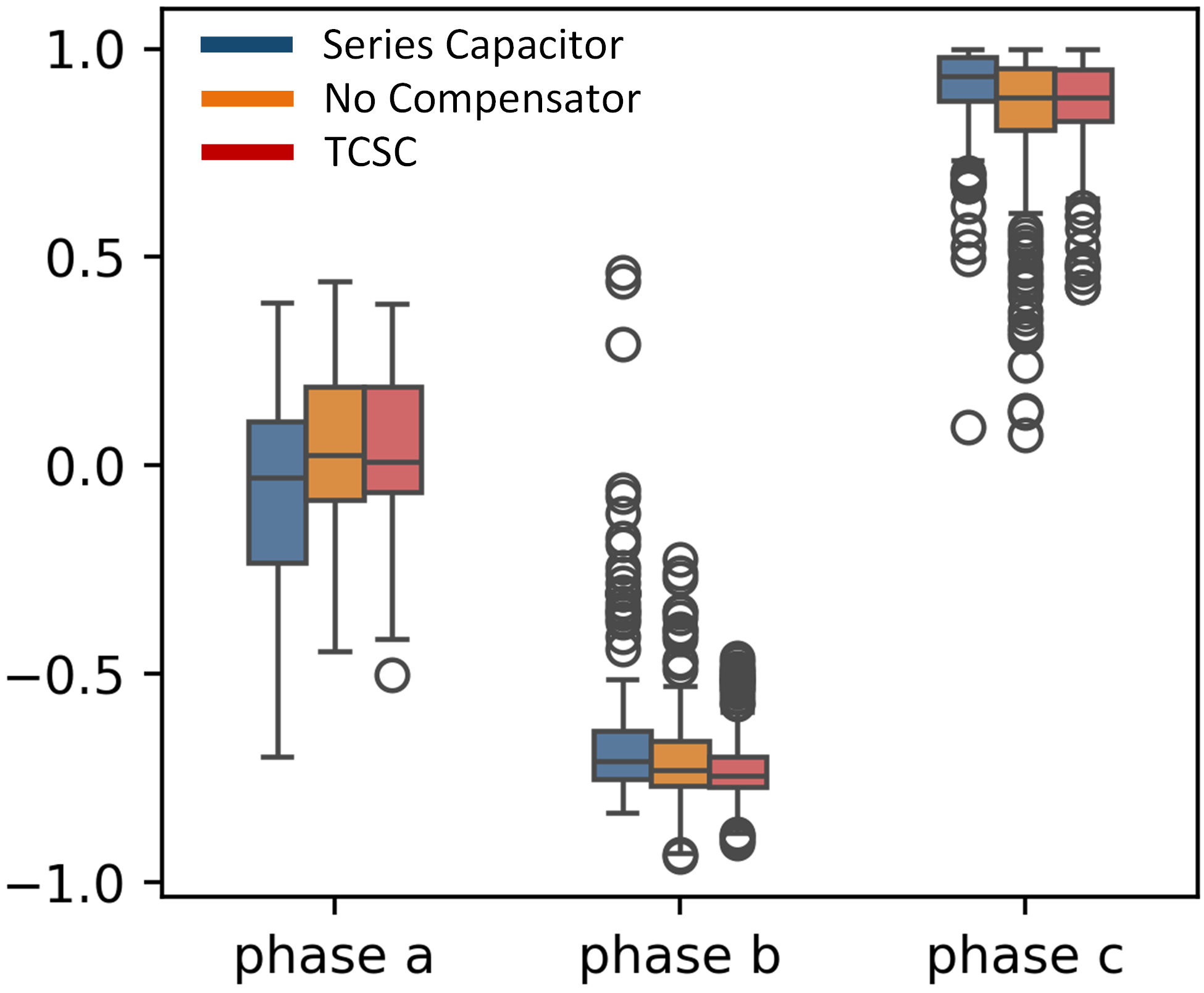} 
	\caption{Spread of CLT features for different compensating devices}\label{tcsc cap}
\end{figure}

\subsection{\textbf{Performance in Presence of TCSC}}
The performance of traditional distance relays may be impacted by TCSCs which are used to improve system stability, increase power transfer capability, and regulate voltage levels in presence of IBR \cite{sauviktcsc21}.  In the 9-bus test system, a TCSC with a capacitor, an inductor, and a metal oxide varistor which handles overvoltages is installed between bus-9 and 4 i.e. at location f6 to provide a maximum compensation of 50\% (Fig. \ref{tcscfig}). By altering the priority (P and Q), fault locations (f2, f5, and f6), fault resistances (0.01, 1, and 10 ohms), fault types (10), and fault inception angles (6) 1080 faults and 2400 switching transients are simulated. The proposed method correctly recognized the transients and faults with a $\bar\eta$ of 96\%.
{Distribution of CLTs for cases with compensation with series capacitor, no compensation, and compensation with TCSC  are shown in Fig. \ref{tcsc cap}. 180 fault cases for each of these 3 scenarios at location f6 are considered. The series capacitor and TCSC boxplots are more skewed than the no compensator boxplots. }


 \begin{figure}[ht]
	\centering
	\includegraphics [width=2.3 in, height= 2.8 in] {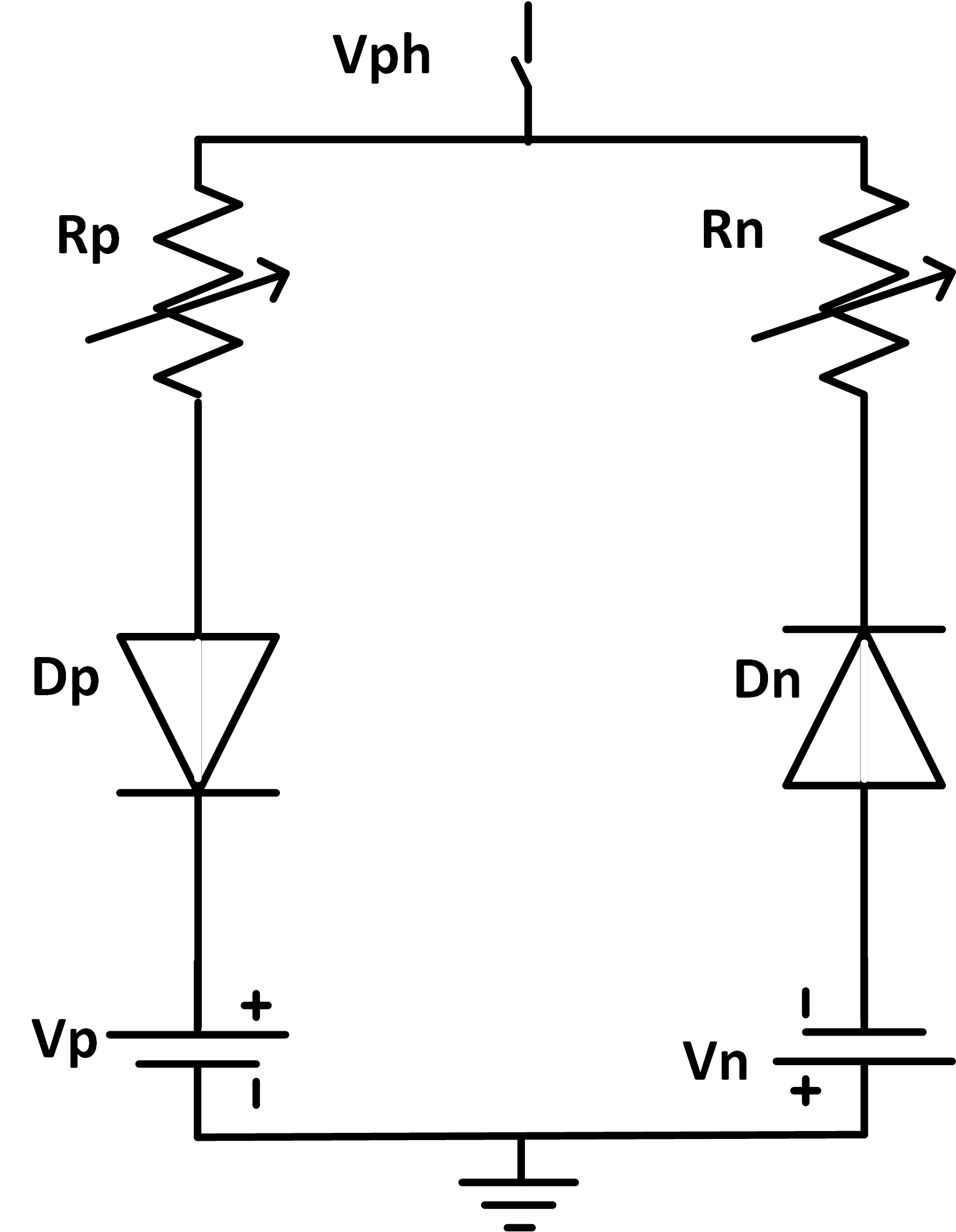} 
	\caption{HIF model}\label{hif}
\end{figure}

 \begin{figure}[ht]
	\centering
	\includegraphics [width=4.5 in, height= 3.9 in] {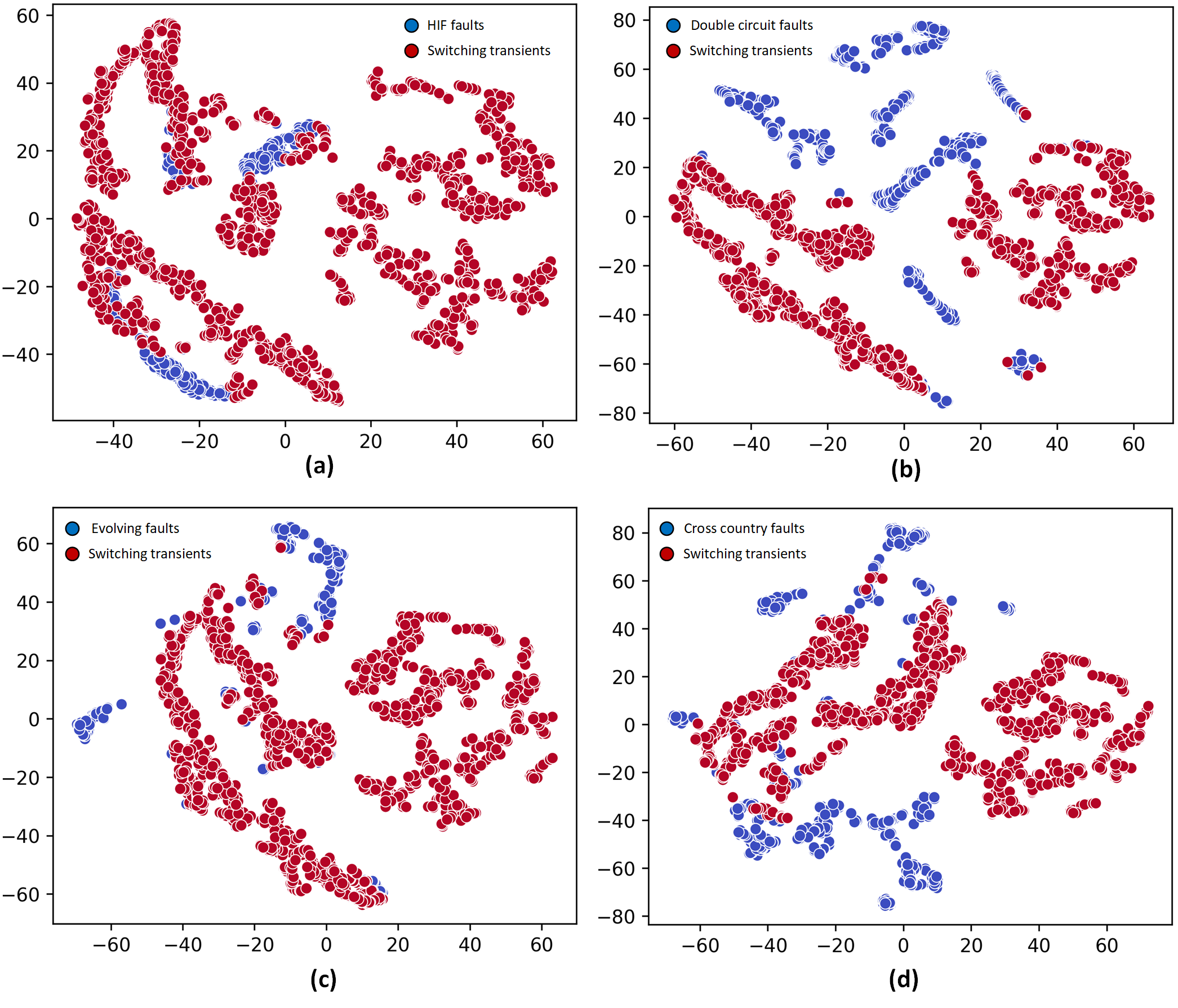} 
	\caption{2D tSNE plots of CLT features for different scenarios}\label{diff scenarios}
\end{figure}

\subsection{\textbf{Performance in Presence of High Impedance Faults}}
Due to relay sensitivity concerns with extremely low-level fault currents and relay design constraints, high impedance faults (HIFs) can be hard to recognize with traditional distance or overcurrent relays \cite{pallavhif}. There are three main approaches to modeling HIFs \cite{hif}. In this work, the circuit-based configuration that uses two diodes and two variable resistors to connect two anti-parallel DC sources is used. The model for HIF employed is illustrated in Fig. \ref{hif}. The dynamic arc is modeled by the two variable resistors, the diodes control the current direction, and the asymmetry in the fault currents is modeled by the varying DC sources.  $Vph > Vn$ in positive half cycle,   $Vph < Vn$ in negative half cycle, and when $Vn < Vph < Vp$ current is zero. 
For the purpose of simulating the 370 HIFs at fault locations f4, f5, f6, f7, and f8 with P and Q priority for 37 fault inception angles, the $lg$ fault in phase-a with fault resistances between 50 ohms and 300 ohms obtained arbitrarily every two milliseconds is taken into consideration. The suggested method successfully distinguished HIFs from other switching transients, achieving a $\bar\eta$ of 96\%. {The t-SNE plot for the 3 CLT features plotted on a 2D plane shows clusters of similar data points, with the distances between these points reflecting their relationships in the original 3-dimensional space. The overlaps in 2D t-SNE plot happens due to the inherent loss of information when reducing dimensions from 3 to 2 (Fig. \ref{diff scenarios}a). }

\subsection{\textbf{Performance in presence of Double Circuit Line}}
The reliability of ground distance relays is threatened by the mutual coupling of double circuit t-lines, demanding additional consideration \cite{double}. So, between buses $3_{PV}$ and 9, a 100 km long double-circuit t-line working at 230 kV and 60 Hz is connected. The suggested scheme recognized the faults from switching transients with a $\bar\eta$ of 99.4\% on 2400 transients and 1260 faults generated in the middle of this t-line in P and Q priority modes for fault resistances (0.01, 1, and 10 ohms), fault types (10), and fault inception angles (21). {Fig. \ref{diff scenarios}b shows the CLT features for faults and switching transients.}


 \begin{figure}[ht]
	\centering
	\captionsetup{textfont=small}
	\includegraphics[width=6.4 in, height=1.4in]{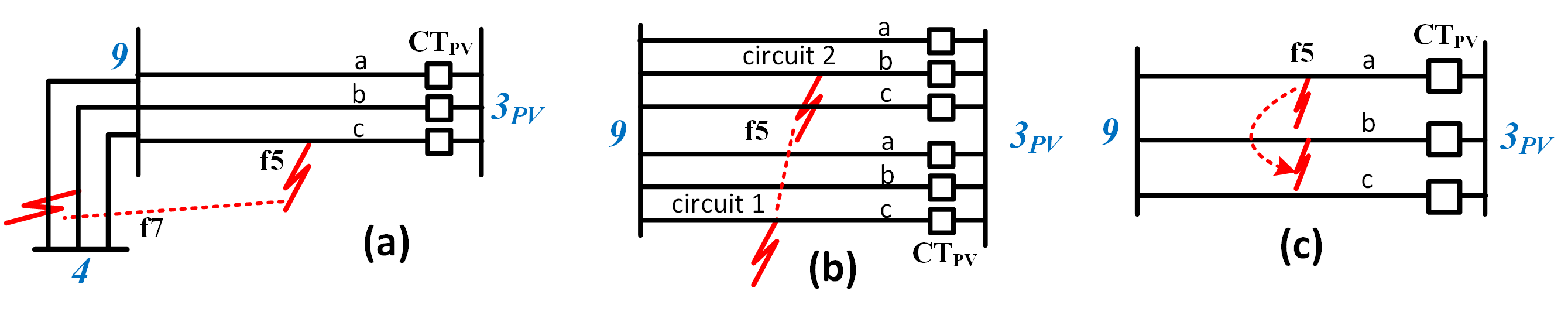}
	\caption{ Cross-country faults:(a) bg at f7 and cg at f5 at 9.0s, (b) cg (circuit 1) and bg (circuit 2) at f5 at 9.0s, (c) evolving fault: ag at 9.0s transformed into abg at 9.008s at f5.}
	\label{faults}
\end{figure}

\subsection{\textbf{Effect of Cross-country Faults and Evolving Faults}}
The effectiveness of the distance relaying scheme is adversely affected by cross-country faults encompassing faults occurring at two separate locations with different or same fault inception time (Fig. \ref{faults}a,b), and evolving faults, which have primary and secondary faults that start at different times but happen at the same place (Fig. \ref{faults}c) \cite{crossevolve}.


The method is assessed across 1188 instances of cross-country faults acquired at P and Q priority by varying fault inception angles (11) and fault resistances (0.01, 1, and 10 ohms). Simultaneous $lg$ faults at location f7 and location f5 (e.g. $bg$  at f7 and $ag$, $bg$, $cg$ at f5)(see Fig. \ref{faults}a) are simulated.
Also, simultaneous $lg$ faults in circuit 1 and circuit 2 (e.g. $cg$ in circuit 1 and $ag$, $bg$, $cg$ in circuit 2) at location f5 (Fig. \ref{faults}b) are simulated. The proposed CLT-based method identified the faults with $\bar\eta$ of 96.6\%.
 
The method is also assessed across 396 instances of evolving faults acquired by varying fault inception angles (11), fault resistances (0.01, 1, and 10 ohms), and P and Q priority. $lg$ faults in one phase get transformed into $llg$ faults involving two phases (e.g. $ag$ to $abg$, $acg$; $bg$ to $abg$, $bcg$; $cg$ to $bcg$, $acg$) at same location f5 (Fig. \ref{faults}c). The faults are found with $\bar\eta$ of 96.5\% by the proposed method.  {Fig. \ref{diff scenarios}c and \ref{diff scenarios}d shows the CLT features for these faults and switching transients.}



\subsection{\textbf{Near-end and Far-end faults}}
Traditional relays might experience malfunctions when facing near-end faults, primarily owing to low voltage and high current magnitudes leading to CT saturation. Additionally, it can be difficult to detect far-end faults since the voltage and current values are both within the usual range.  
By altering the fault inception angles, fault resistances, priority, and fault types, 720 faults were produced at locations f5 and f3, which were used as the far-end and near-end faults, respectively. RF distinguishes these faults with $\bar\eta$ of 99.3\%.

\begin{figure}[ht]
	\centering
	\captionsetup{justification=centering,textfont=small}
	\includegraphics[width=4.3 in, height=4.1 in]{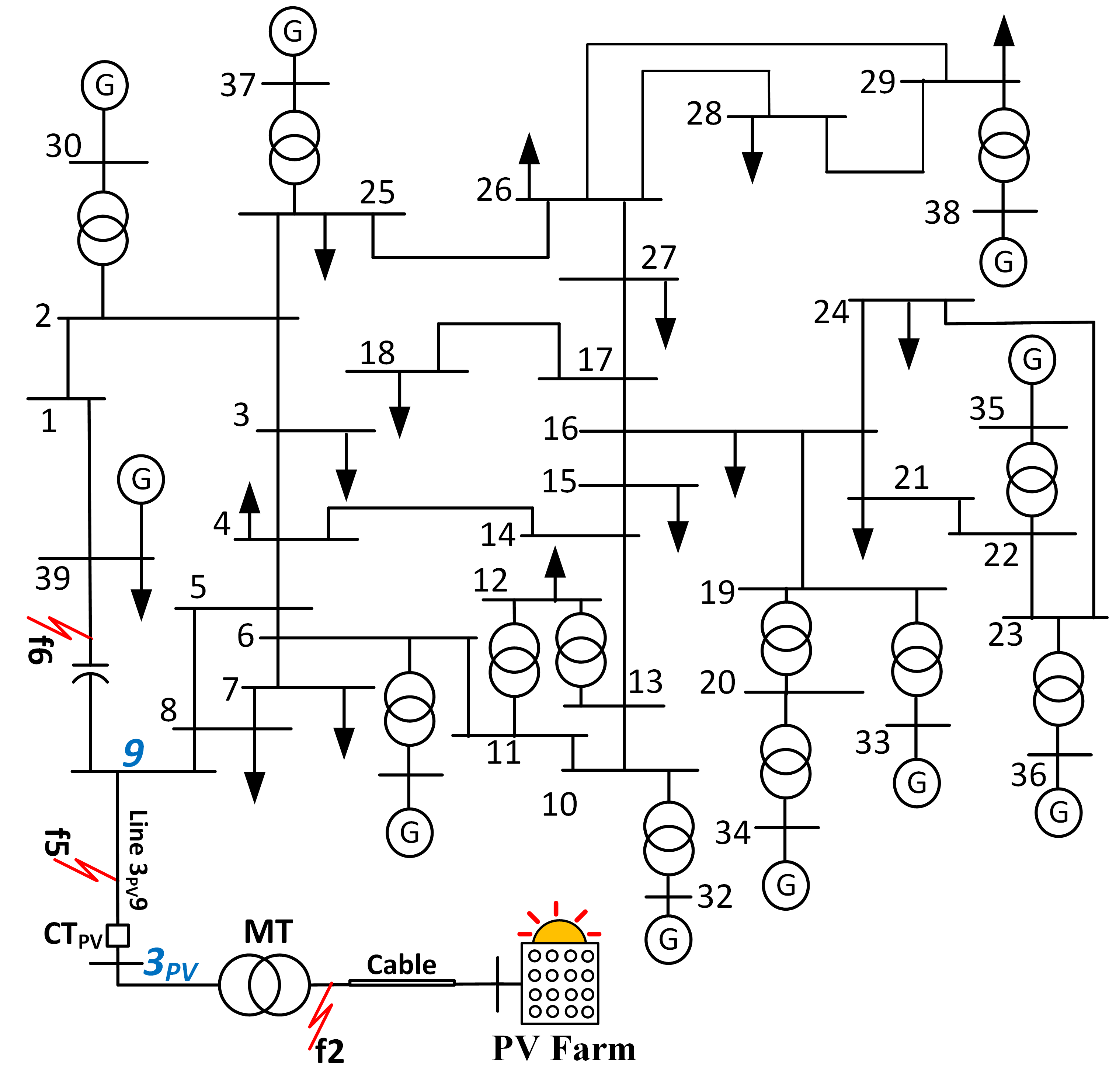}
	\caption{IEEE 39-bus with PV at bus-9}
	\label{39busSYSTEM}
\end{figure}

\subsection{\textbf{Effect of change in system}}
The IEEE 39-bus system is also used to assess the viability of the proposed scheme (Fig.\ref{39busSYSTEM}).
The 39-bus system is also served by the CLT-based fault detection, which has a $\bar\eta$ of 98.0\% on 1080 faults and 2400 non-fault cases. The priority (P and Q), fault locations (f2, f5, and f6), fault resistances (0.01, 1, and 10 ohms), fault types (10), and fault inception angles (6) are all altered to simulate the faults.

\begin{figure}[htpb]
	\centering
	\captionsetup{justification=centering,textfont=small}
	\includegraphics[width=4.1 in, height=3.1 in]{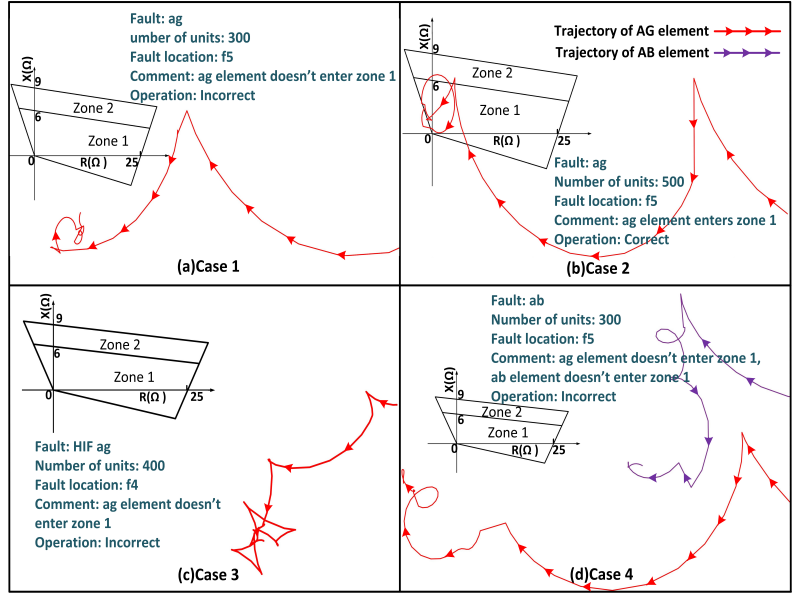}
	\caption{Impedance trajectories for AG and AB components of the distance relay at $CT_{PV}$ for faults in zone 1}
	\label{pvmaloperate}
\end{figure}

\begin{table}[ht]
\centering
\scriptsize
\renewcommand{\arraystretch}{1.3}
\setlength{\tabcolsep}{4 pt}
\caption{Comparison of recently published articles.}\label{comparison}
\begin{tabular}{|>{\itshape}l|c|c|c|c|c|} \hline
\rowcolor[rgb]{0.91,0.91,0.91} Reference                                                                   & \cite{saber22} &\cite{AKTER2022}  &  \cite{GHORBANI2023}& \cite{sauviktcsc21} & \begin{tabular}[c]{@{}l@{}}Proposed \\ method\end{tabular} \\ \hline
Technique used                                                                      & \begin{tabular}[c]{@{}l@{}}signed \\correlation \end{tabular}& impedance & \begin{tabular}[c]{@{}l@{}}+tive seq.\\network\end{tabular} & EMD,RF & \begin{tabular}[c]{@{}l@{}}CLT,\\Fuzzy,RF \end{tabular}                                             \\ \hline
Signals used                                                                &  $i$& $ v$ \& $i$ & $ v$ \& $i$ & $i$  &  $i$                                                     \\ \hline
\begin{tabular}[c]{@{}l@{}}Single or double end\end{tabular}                & double & single &single &  single& single                                                     \\ \hline
\begin{tabular}[c]{@{}l@{}}System freq.(Hz), Samp. freq.(kHz)\end{tabular} & 60,1.2 & 60,1 &  60,1.2&  50,1& 60,7.84                                                   \\ \hline
Model includes                                                           & WF & PV &  PV&  WF& PV                                                  \\ \hline
FACTS used                                                                  & - &  -& - &  TCSC& TCSC                                                 \\ \hline
Time delay (ms)                                                              & \textasciitilde16 & 16.67 &  -&  8& 18.5                                                        \\ \hline
Effect of HIF                                                               & - & - & yes & yes & yes                                                        \\ \hline
Effect of Noise                                                             & yes & yes & yes & yes & yes                                                        \\ \hline
\begin{tabular}[c]{@{}l@{}}Effect of Double ckt. lines\end{tabular}       & - & - & - & - & yes                                                        \\\hline
\begin{tabular}[c]{@{}l@{}}Effect of farm capacity\end{tabular}    &yes & - & - & - & yes                                                        \\\hline
\begin{tabular}[c]{@{}l@{}}Effect of cross-country faults\end{tabular}    &-  & - & - & - & yes                                                        \\\hline
\begin{tabular}[c]{@{}l@{}}Effect of evolving faults\end{tabular}         & - & yes & yes & - & yes                                                        \\\hline
\begin{tabular}[c]{@{}l@{}}Effect of Sampling freq.\end{tabular}         & - & - & - & - & yes                                                        \\\hline
\begin{tabular}[c]{@{}l@{}}Effect of data window\end{tabular}             & - & - & - & yes & yes                                                        \\\hline
\begin{tabular}[c]{@{}l@{}}Effect of CT saturation\end{tabular}           & - & yes & yes & yes & yes                                                        \\\hline
\begin{tabular}[c]{@{}l@{}}Effect of load switching\end{tabular}          & yes & yes & - & yes & yes                                                        \\\hline
\begin{tabular}[c]{@{}l@{}}Effect of capacitor switching\end{tabular}     & yes & - & - & yes & yes                                                       \\ \hline
\begin{tabular}[c]{@{}l@{}}Effect of near-end faults\end{tabular}     & - &  yes & - & yes & yes                                                       \\ \hline
\end{tabular}
\end{table}

\section{Comparative Evaluation}
This section illustrates the conduct of the conventional distance relay connected to a PV farm, while also highlighting the lack of comprehensiveness of results from recent works to establish the effectiveness of the proposed algorithm in this article.

The impedance plane associated with traditional distance relay $3_{PV}$9, near bus $3_{PV}$, demonstrates the fault behavior. It's evident that the distance relay exhibits unreliable responses (Fig. \ref{pvmaloperate}). For instance, during an $ag$ fault occurring at location f5 (refer Fig. \ref{9busPV}) within zone 1, the relay remains inactive in case 1 (PV units = 300, Rf = 10$\Omega$, priority = P) but it triggers in case 2 (PV units = 500, Rf = 1$\Omega$, priority = P). It remains dormant for a high impedance $ag$ fault at location f4 (zone 1) in case 3 (PV units = 400, priority = Q). Also, it stays inactive for an $ab$ fault at location f5 in case 4 (PV units = 300, Rf = 10$\Omega$, priority = P).  The procedure for obtaining the impedance trajectory is described in \cite{pallavauto}.

Additionally, Table \ref{comparison} compares current publications concentrating on criteria, with particular emphasis on the impact of different transient conditions on the proposed techniques. This evaluation aims to assess the comprehensiveness of these proposed techniques. While the methodologies have produced favorable results under some particular scenarios, performance analysis in many significant situations appears to be lacking. The aforementioned publications also touch on the proposed algorithms' execution times. For double-ended protection, the communication paths between the two terminals can introduce a time delay of up to 6.87ms, according to IEEE standard \cite{standard2}. The proposed technique being single-ended is free from this delay. The intelligent protection system based on CLT depends on data preparation and inference time. The event detector takes 0.001ms and the calculation of CLT takes 0.01ms. The fuzzy and the ML systems work in parallel taking 1.8ms and 0.5ms respectively to test new inputs.  Hence, net running time considering 1-cycle data = 16.67ms + 1.8ms + 0.01ms + 0.001ms = 18.5ms.

\section{CONCLUSION}
The attributes of dependability, security, selectivity, robustness, and speed should be present in a protective system.
However, the dependability for internal faults and security for external faults and other transients may get challenged in the case of transmission lines connected to bulk PV farms. 
 The suggested combined linear trend-based hybrid intelligent protection method is dependable for internal faults, cross-country faults, evolving faults, double circuit line faults; and responsive to low current levels in high impedance faults. It provides security for capacitor and load-switching, and external fault with CT saturation events.
It locates the faults correctly and is selective. It is resilient to changes in data window size,  capacity of the PV, noise in measurements, change in the test system, and the presence of TCSC. However, the performance of the model is impacted by the sampling rate of the phase currents.
The combined linear trend-based intelligent approach has been successfully evaluated in the IEEE-9 bus system across a wide range of fault parameters and operating scenario variations. It has been found that the linear combined trend-based decision-making system provides a thorough protection for transmission lines connected to bulk photovoltaic farms while being dependable, secure, accurate, and quick. Furthermore, it solely utilizes locally measured data, eliminating the requirement for a remote-end communication device. The proposed method can be applied in real time using edge computing for power system protection with careful planning, cybersecurity measures, and integration with existing infrastructure, thereby enhancing the speed, reliability, and scalability of protection functions.

\bibliographystyle{elsarticle-num} 
\small{
\bibliography{references}}

\begin{thebibliography}{10}
\expandafter\ifx\csname url\endcsname\relax
  \def\url#1{\texttt{#1}}\fi
\expandafter\ifx\csname urlprefix\endcsname\relax\def\urlprefix{URL }\fi
\expandafter\ifx\csname href\endcsname\relax
  \def\href#1#2{#2} \def\path#1{#1}\fi

\bibitem{bansal}
R.~Bansal, Power System Protection in Smart Grid Environment, CRC Press, New
  York, USA, 2019.
\newblock \href {https://doi.org/10.1201/9780429401756}
  {\path{doi:10.1201/9780429401756}}.

\bibitem{singh2018}
A.~K. Singh, I.~Hussain, B.~Singh, Double-stage three-phase grid-integrated
  solar {PV} system with fast zero attracting normalized least mean fourth
  based adaptive control, IEEE Trans. Ind. Electron. 65~(5) (2018) 3921--3931.
\newblock \href {https://doi.org/10.1109/TIE.2017.2758750}
  {\path{doi:10.1109/TIE.2017.2758750}}.

\bibitem{sikander}
S.~Singh, P.~K. Nayak, S.~Sarangi, S.~Biswas, Improved protection scheme for
  high voltage transmission lines connecting large-scale solar {PV} plants, in:
  22nd National Power Systems Conference (NPSC), 2022, pp. 118--123.
\newblock \href {https://doi.org/10.1109/NPSC57038.2022.10069457}
  {\path{doi:10.1109/NPSC57038.2022.10069457}}.

\bibitem{Banaiemoqadam2020}
A.~Banaiemoqadam, A.~Hooshyar, M.~A. Azzouz, A control-based solution for
  distance protection of lines connected to converter-interfaced sources during
  asymmetrical faults, IEEE Trans. Power Del. 35~(3) (2020) 1455--1466.
\newblock \href {https://doi.org/10.1109/TPWRD.2019.2946757}
  {\path{doi:10.1109/TPWRD.2019.2946757}}.

\bibitem{liang}
Y.~Liang, W.~Li, W.~Zha, Adaptive mho characteristic-based distance protection
  for lines emanating from photovoltaic power plants under unbalanced faults,
  IEEE Syst. J. 15~(3) (2021) 3506--3516.
\newblock \href {https://doi.org/10.1109/JSYST.2020.3015225}
  {\path{doi:10.1109/JSYST.2020.3015225}}.

\bibitem{ritwik}
R.~Chowdhury, N.~Fischer, Transmission line protection for systems with
  inverter-based resources – part i: Problems, IEEE Trans. Power Del. 36~(4)
  (2021) 2416--2425.
\newblock \href {https://doi.org/10.1109/TPWRD.2020.3019990}
  {\path{doi:10.1109/TPWRD.2020.3019990}}.

\bibitem{kou2020}
G.~Kou, J.~Jordan, B.~Cockerham, R.~Patterson, P.~VanSant, Negative-sequence
  current injection of transmission solar farms, IEEE Trans. Power Del. 35~(6)
  (2020) 2740--2743.
\newblock \href {https://doi.org/10.1109/TPWRD.2020.3014783}
  {\path{doi:10.1109/TPWRD.2020.3014783}}.

\bibitem{haddadi2021}
A.~Haddadi, M.~Zhao, I.~Kocar, U.~Karaagac, K.~W. Chan, E.~Farantatos, Impact
  of inverter-based resources on negative sequence quantities-based protection
  elements, IEEE Trans. Power Del. 36~(1) (2021) 289--298.
\newblock \href {https://doi.org/10.1109/TPWRD.2020.2978075}
  {\path{doi:10.1109/TPWRD.2020.2978075}}.

\bibitem{pvqiang}
Q.~Huang, K.~Li, Y.~Li, R.~Fan, A.~Wang, G.~Zhang, Y.~Sun, Adaptability
  analysis of traditional differential protection applied to lines connected to
  {PV}, in: IEEE 5th Advanced Information Management, Communicates, Electronic
  and Automation Control Conference (IMCEC), Vol.~5, 2022, pp. 1819--1822.
\newblock \href {https://doi.org/10.1109/IMCEC55388.2022.10019865}
  {\path{doi:10.1109/IMCEC55388.2022.10019865}}.

\bibitem{yang}
Z.~{Yang}, K.~{Jia}, Y.~{Fang}, Z.~{Zhu}, B.~{Yang}, T.~{Bi}, High-frequency
  fault component-based distance protection for large renewable power plants,
  IEEE Trans. Power Electron. 35~(10) (2020) 10352--10362.
\newblock \href {https://doi.org/10.1109/TPEL.2020.2978266}
  {\path{doi:10.1109/TPEL.2020.2978266}}.

\bibitem{chen}
B.~{Chen}, A.~{Shrestha}, F.~A. {Ituzaro}, N.~{Fischer}, Addressing protection
  challenges associated with type 3 and type 4 wind turbine generators, in:
  68th Annual Conference for Protective Relay Engineers, 2015, pp. 335--344.
\newblock \href {https://doi.org/10.1109/CPRE.2015.7102177}
  {\path{doi:10.1109/CPRE.2015.7102177}}.

\bibitem{kumar}
D.~S. Kumar, Impact analysis of distributed generators on the protection and
  stability of power systems, Ph.D. thesis, National University of Singapore
  (2017).

\bibitem{hosiyar2}
{A. {Hooshyar}, M. A. {Azzouz}, and E. F. {El-Saadany}}, Three-phase fault
  direction identification for distribution systems with {DFIG}-based wind
  {DG}, IEEE Trans. Sustain. Energy 5~(3) (2014) 747--756.
\newblock \href {https://doi.org/10.1109/TSTE.2014.2298466}
  {\path{doi:10.1109/TSTE.2014.2298466}}.

\bibitem{kavi}
M.~Kavi, Y.~Mishra, M.~Vilathgamuwa, Challenges in high impedance fault
  detection due to increasing penetration of photovoltaics in radial
  distribution feeder, in: IEEE Power \& Energy Society General Meeting, 2017,
  pp. 1--5.
\newblock \href {https://doi.org/10.1109/PESGM.2017.8274658}
  {\path{doi:10.1109/PESGM.2017.8274658}}.

\bibitem{fang2019}
Y.~Fang, K.~Jia, Z.~Yang, Y.~Li, T.~Bi, Impact of inverter-interfaced renewable
  energy generators on distance protection and an improved scheme, IEEE Trans.
  Ind. Electron. 66~(9) (2019) 7078--7088.
\newblock \href {https://doi.org/10.1109/TIE.2018.2873521}
  {\path{doi:10.1109/TIE.2018.2873521}}.

\bibitem{pallavauto}
P.~K. Bera, V.~Kumar, S.~R. Pani, O.~P. Malik, Autoregressive coefficients
  based intelligent protection of transmission lines connected to type-3 wind
  farms, IEEE Trans. Power Del. 39~(1) (2024) 71--82.
\newblock \href {https://doi.org/10.1109/TPWRD.2023.3321844}
  {\path{doi:10.1109/TPWRD.2023.3321844}}.

\bibitem{saber22}
A.~Saber, M.~Shaaban, H.~Zeineldin, A new differential protection algorithm for
  transmission lines connected to large-scale wind farms, Int. J. Elect. Power
  Energy Syst. 141 (2022) 108220.
\newblock \href {https://doi.org/https://doi.org/10.1016/j.ijepes.2022.108220}
  {\path{doi:https://doi.org/10.1016/j.ijepes.2022.108220}}.

\bibitem{JIA2018}
K.~Jia, Y.~Li, Y.~Fang, L.~Zheng, T.~Bi, Q.~Yang, Transient current similarity
  based protection for wind farm transmission lines, Applied Energy 225 (2018)
  42--51.
\newblock \href
  {https://doi.org/https://doi.org/10.1016/j.apenergy.2018.05.012}
  {\path{doi:https://doi.org/10.1016/j.apenergy.2018.05.012}}.

\bibitem{GHORBANI2023}
A.~Ghorbani, H.~Mehrjerdi, Distance protection with fault resistance
  compensation for lines connected to {PV} plant, Int. J. Elect. Power Energy
  Syst. 148 (2023) 108976.
\newblock \href {https://doi.org/https://doi.org/10.1016/j.ijepes.2023.108976}
  {\path{doi:https://doi.org/10.1016/j.ijepes.2023.108976}}.

\bibitem{AKTER2022}
S.~Akter, S.~Biswal, A.~R. Adly, A.~Chakraborti, B.~K. Roy, P.~Das, A.~Y.
  Abdelaziz, Impedance based directional relaying for smart power networks
  integrating with converter interfaced photovoltaic plants, Electr. Power
  Syst. Res. 213 (2022) 108711.
\newblock \href {https://doi.org/https://doi.org/10.1016/j.epsr.2022.108711}
  {\path{doi:https://doi.org/10.1016/j.epsr.2022.108711}}.

\bibitem{omar2018}
O.~Noureldeen, I.~Hamdan, Design of robust intelligent protection technique for
  large-scale grid-connected wind farm, Protection and Control of Modern Power
  Systems 3~(17) (2018) 1--13.
\newblock \href {https://doi.org/10.1186/s41601-018-0090-4}
  {\path{doi:10.1186/s41601-018-0090-4}}.

\bibitem{sauviktcsc21}
S.~Biswas, P.~K. Nayak, A new approach for protecting tcsc compensated
  transmission lines connected to {DFIG}-based wind farm, IEEE Trans. Ind.
  Inform. 17~(8) (2021) 5282--5291.
\newblock \href {https://doi.org/10.1109/TII.2020.3029201}
  {\path{doi:10.1109/TII.2020.3029201}}.

\bibitem{pvsvm}
Y.~B. Yoldas, R.~Yumurtacı, {Improvement of Distance Protection with SVM on
  {PV}-Fed Transmission Lines in Infeed Conditions}, Energies 16~(6) (2023)
  1--18.

\bibitem{mdpi_pvwf}
K.~Al~Kharusi, A.~El~Haffar, M.~Mesbah, Fault detection and classification in
  transmission lines connected to inverter-based generators using machine
  learning, Energies 15~(15) (2022).
\newblock \href {https://doi.org/10.3390/en15155475}
  {\path{doi:10.3390/en15155475}}.

\bibitem{CHOWDHURY}
A.~Chowdhury, S.~Paladhi, A.~K. Pradhan, Local positive sequence component
  based protection of series compensated parallel lines connecting solar
  photovoltaic plants, Electric Power Systems Research 225 (2023) 109811.
\newblock \href {https://doi.org/https://doi.org/10.1016/j.epsr.2023.109811}
  {\path{doi:https://doi.org/10.1016/j.epsr.2023.109811}}.

\bibitem{datapv}
P.~Bera, S.~Pani, C.~Isik, R.~Bansal, Transients in transmission lines
  connected to photovoltaic farms (dataset) (2023).
\newblock \href {https://doi.org/https://dx.doi.org/10.21227/scrs-bs27}
  {\path{doi:https://dx.doi.org/10.21227/scrs-bs27}}.

\bibitem{pscad}
E.~Muljadi, M.~Singh, V.~Gevorgian, User guide for {PV} dynamic model
  simulation written on pscad platform, Tech. rep., National Renewable Energy
  Lab.(NREL), Golden, CO (United States) (2014).

\bibitem{greywolf}
S.~Mirjalili, S.~M. Mirjalili, A.~Lewis, Grey wolf optimizer, Advances in
  Engineering Software 69 (2014) 46--61.
\newblock \href
  {https://doi.org/https://doi.org/10.1016/j.advengsoft.2013.12.007}
  {\path{doi:https://doi.org/10.1016/j.advengsoft.2013.12.007}}.

\bibitem{pedes}
S.~R. Pani, P.~K. Bera, V.~Kumar, Detection and classification of internal
  faults in power transformers using tree based classifiers, in: 2020 IEEE
  International Conference on Power Electronics, Drives and Energy Systems
  (PEDES), 2020, pp. 1--6.
\newblock \href {https://doi.org/10.1109/PEDES49360.2020.9379641}
  {\path{doi:10.1109/PEDES49360.2020.9379641}}.

\bibitem{systempallav}
P.~K. Bera, C.~Isik, V.~Kumar, Discrimination of internal faults and other
  transients in an interconnected system with power transformers and phase
  angle regulators, IEEE Syst. J. 15~(3) (2021) 3450--3461.
\newblock \href {https://doi.org/10.1109/JSYST.2020.3009203}
  {\path{doi:10.1109/JSYST.2020.3009203}}.

\bibitem{powerswing}
P.~K. Bera, C.~Isik, Identification of stable and unstable power swings using
  pattern recognition, in: 2021 IEEE Green Technologies Conference (GreenTech),
  2021, pp. 286--291.
\newblock \href {https://doi.org/10.1109/GreenTech48523.2021.00053}
  {\path{doi:10.1109/GreenTech48523.2021.00053}}.

\bibitem{tsfresh}
M.~Christ, N.~Braun, J.~Neuffer, A.~W. Kempa-Liehr,
  \href{https://www.sciencedirect.com/science/article/pii/S0925231218304843}{Time
  series feature extraction on basis of scalable hypothesis tests (tsfresh –
  a python package)}, Neurocomputing 307 (2018) 72--77.
\newblock \href {https://doi.org/https://doi.org/10.1016/j.neucom.2018.03.067}
  {\path{doi:https://doi.org/10.1016/j.neucom.2018.03.067}}.
\newline\urlprefix\url{https://www.sciencedirect.com/science/article/pii/S0925231218304843}

\bibitem{smote}
N.~V. Chawla, et~al., {SMOTE}: Synthetic minority over-sampling technique, J.
  Artif. Int. Res. 16~(1) (2002) 321–357.

\bibitem{tsne}
L.~van~der Maaten, G.~Hinton, Visualizing data using {t-SNE}, Journal of
  Machine Learning Research 9 (2008) 2579--2605.

\bibitem{cvt}
D.~Costello, K.~Zimmerman, {CVT} transients revisited — distance, directional
  overcurrent, and communications-assisted tripping concerns, in: 2012 65th
  Annual Conference for Protective Relay Engineers, 2012, pp. 73--84.
\newblock \href {https://doi.org/10.1109/CPRE.2012.6201222}
  {\path{doi:10.1109/CPRE.2012.6201222}}.

\bibitem{pallavhif}
P.~K. Bera, V.~Kumar, S.~R. Pani, V.~Bargate, Detection of high impedance
  faults in microgrids using machine learning, in: IEEE Green Energy and Smart
  System Systems (IGESSC), 2022, pp. 1--5.
\newblock \href {https://doi.org/10.1109/IGESSC55810.2022.9955330}
  {\path{doi:10.1109/IGESSC55810.2022.9955330}}.

\bibitem{hif}
Q.~Cui, K.~El-Arroudi, Y.~Weng, A feature selection method for high impedance
  fault detection, IEEE Trans. Power Del. 34~(3) (2019) 1203--1215.
\newblock \href {https://doi.org/10.1109/TPWRD.2019.2901634}
  {\path{doi:10.1109/TPWRD.2019.2901634}}.

\bibitem{double}
Y.~Zhang, J.~Liang, Z.~Yun, X.~Dong, A new fault-location algorithm for
  series-compensated double-circuit transmission lines based on the distributed
  parameter model, IEEE Trans. Power Del. 32~(6) (2017) 2398--2407.
\newblock \href {https://doi.org/10.1109/TPWRD.2016.2626476}
  {\path{doi:10.1109/TPWRD.2016.2626476}}.

\bibitem{crossevolve}
V.~Ashok, A.~Yadav, A.~Y. Abdelaziz, Modwt-based fault detection and
  classification scheme for cross-country and evolving faults, Electr. Power
  Syst. Res. 175 (2019) 105897.
\newblock \href {https://doi.org/https://doi.org/10.1016/j.epsr.2019.105897}
  {\path{doi:https://doi.org/10.1016/j.epsr.2019.105897}}.

\bibitem{standard2}
{IEEE} {G}uide for {A}pplication of {D}igital line {C}urrent {D}ifferential
  {R}elays using {D}igital {C}ommunication, IEEE Std C37.243-2015 (2015)
  1--72\href {https://doi.org/10.1109/IEEESTD.2015.7181615}
  {\path{doi:10.1109/IEEESTD.2015.7181615}}.

\end{thebibliography}

\end{document}